\documentclass[aps, pra, amsmath, reprint, superscriptaddress]{revtex4-2}

\usepackage{times, newtxmath, graphicx, textgreek, tikz, placeins, xcolor, siunitx}
\usepackage[version = 4]{mhchem}
\usepackage[final,stretch = 10, shrink = 10]{microtype}
\usepackage{hyperref}
\usepackage[capitalize]{cleveref}
\usepackage{mathtools}

\hypersetup{colorlinks=true,linkcolor=red,urlcolor=blue,anchorcolor=blue,citecolor=blue}
\allowdisplaybreaks[1]
\graphicspath{{/Images/}}

\definecolor{darkBlue}{RGB}{0,0,127}
\definecolor{darkRed}{RGB}{192,0,0}

\crefname{section}{Sec.}{Secs.}
\Crefname{section}{Section}{Sections}

\newcommand{\Nottingham}{School of Physics and Astronomy, University of Nottingham, Nottingham, NG7 2RD, United Kingdom}
\newcommand{\Theoretical}{Centre for the Mathematics and Theoretical Physics of Quantum Non-Equilibrium Systems, University of Nottingham, Nottingham NG7 2RD, United Kingdom}
\newcommand{\bosonic}{\omega}
\newcommand{\qubit}{\Omega}
\newcommand{\coupling}{g}
\newcommand{\asymmetry}{\epsilon}
\newcommand{\Gfunc}{\mathcal{G}}

\newcommand{\ex}[1]{\ensuremath{\mathrm{e}^{#1}}}

\DeclareMathOperator{\tr}{tr}

\begin{document}

\title{Tripartite quantum Rabi model with trapped Rydberg ions}
\author{Thomas~J.\ \surname{Hamlyn}}
\affiliation{\Nottingham}
\affiliation{\Theoretical}
\author{Chi \surname{Zhang}}
\affiliation{Division of Physics, Mathematics, and Astronomy, California Institute of Technology, Pasadena, CA 91125, USA}
\author{Igor \surname{Lesanovsky}}
\affiliation{Institut f\"{u}r Theoretische Physik, Universit\"{a}t T\"{u}bingen, Auf der Morgenstelle 14, 72076 T\"{u}bingen, Germany}
\affiliation{\Nottingham}
\affiliation{\Theoretical}
\author{Weibin \surname{Li}}
\affiliation{\Nottingham}
\affiliation{\Theoretical}
	
\begin{abstract}
We investigate a tripartite quantum Rabi model (TQRM) wherein a bosonic mode concurrently couples to two spin-$1/2$ particles through a spin-spin interaction, resulting in a spin-spin-boson coupling -- a departure from conventional quantum Rabi models featuring bipartite spin-boson couplings. The symmetries of the TQRM depend on the detuning parameter, representing the energy difference between the spin states. At zero detuning a parity symmetry renders the TQRM reducible to a quantum Rabi model. A subradiant to superradiant transition in the groundstate is predicted as the tripartite coupling strength increases.
For non-zero detuning the total spin emerges as the sole conserved quantity in the TQRM. It is found that superradiance prevails in the groundstate as long as the tripartite coupling remains non-zero. We derive the Braak $\Gfunc$-function of the TQRM analytically, with which the eigenspectra are obtained. The TQRM can be realized in a viable trapped Rydberg ion quantum simulator, where the required tripartite couplings and single body interactions in the TQRM are naturally present. Our study opens opportunities to explore and create novel correlations and entanglement in the spin and motional degrees of freedoms with the TQRM.
\end{abstract}
\maketitle

\section{Introduction}
The~quantum Rabi model (QRM) consists of a spin-$1/2$ particle coupled to a bosonic degree of freedom~\cite{Rabi_1936}. Despite its simplicity the QRM exhibits rich physics~\cite{Scully_1997}, and finds applications in, e.g., benchmarking quantum computers~\cite{Burger_2022}, verifying the existence of supersymmetric quantum mechanics~\cite{Hirokawa_2015, Cai_2022}, investigating $\mathcal{PT}$-symmetry breaking~\cite{Lee_2015, Lu_2022}, and~the generation of nonclassical states~\cite{Ashhab_2010, Leroux_2017}. An important feature of the QRM is that the validity of the rotating wave approximation breaks down as a result of the strong spin-boson coupling~\cite{Kockum_2019, Le_Boite_2020,Irish_2022}. This causes difficulties in obtaining the spectrum of the QRM analytically. Such a task was not accomplished until 2011 by \citet{Braak_2011} through the $\Gfunc$-function method, although the isolated doubly degenerate energies that represent the exceptional spectrum were analytically determined decades earlier~\cite{Judd_1979}. After this achievement, many studies have focused on understanding the mathematical structure and improving the numerical stability and performance of the $\Gfunc$-function~\cite{Maciejewski_2012, Braak_2013, Moroz_2013, Zhong_2013, Zhong_2014, Xie_2017}. When the coupling is strong the groundstate of the QRM exhibits a superradiant-like phase transition, where the population of the boson becomes sizable when increasing the coupling above a critical value~\cite{Hwang_2015, Felicetti_2020, Wang_2020, Yang_2023}. Motivated by the rich physics, the QRM has been experimentally realized with various physical systems~\cite{Xie_2017}, including~trapped ions~\cite{Lv_2018, Cai_2021}, cavity and circuit QED~\cite{Raimond_2001, Wallraff_2004, Niemczyk_2010, Xiang_2013, Gu_2017, Rose_2017, Blais_2021}, cavity optomechanics~\cite{Dare_2023}, and~quantum dots~\cite{Stievater_2001, Englund_2007}. Furthermore, superradiant transition has been experimentally demonstrated with a single trapped ion~\cite{Cai_2021}. 
	
Modified QRMs have been extensively explored within the literature~\cite{Xie_2017}. The~first type of extension concerns the modification of the bosonic operators. For example two photon QRMs, where the spin couples to the bosonic mode via a two photon process, have attracted a lot of attention~\cite{Stufler_2006, Zhong_2014, Duan_2016, Puebla_2017, Felicetti_2018, Cheng_2018, Chen_2022}. The symmetry extends from a $\mathbb{Z}_2$ group in the QRM to a~$\mathbb{Z}_4$ group in the two photon QRM~\cite{Duan_2016}. Despite the change of symmetry the spectra of the two photon QRM can still be found analytically~\cite{Travenec_2012,  Chen_2012}. Another way to extend the QRM is to include more spins in the model, such as two qubit  QRMs (TQQRMs)~\cite{Peng_2012, Chilingaryan_2013, Peng_2013, Wang_2014, Peng_2014, Rodriguez-Lara_2014, Duan_2015, Mao_2015, Sun_2020, Yan_2021,Grimaudo_2023, Grimaudo_2023b}. Such extension leads to the Dicke model when number of spins is large~\cite{kirton_introduction_2019}. However, all of these existing studies have focused on bipartite couplings which couple individual spins with bosonic modes.
	
In~this article we study an exotic tripartite quantum Rabi model (TQRM) where two identical spins couple to a monochromatic bosonic mode simultaneously through an Ising spin-spin interaction. To the best of our knowledge, such extension has not been studied before. Three- and multi-body interactions have attracted a broad interest in the study of nuclear, atomic and many-body physics ~\cite{hammerColloquiumThreebodyForces2013}. Recent experimental and theoretical studies have realized multi-body interactions with trapped ions~\cite{PRXQuantum.4.030311}, neutral atoms~\cite{goban_emergence_2018}, and superconducting circuits~\cite{PhysRevX.10.011011,PhysRevA.107.013715}. A key feature of the TQRM is that the constituents of the TQRM are a bosonic mode and two identical spins. We show that the TQRM can be realized with two trapped ions in their highly excited Rydberg electronic states~\cite{higgins_single_2017,Zhang_2020a}. In a linear Paul trap long-range dipole-dipole interactions between Rydberg ions couple to the breathing mode of a two-ion crystal~\cite{James_1998}, leading to the novel tripartite coupling between the two spins and bosonic mode (crystal phonons). The TQRM is achieved through modulating the Rydberg excitation and phonon-ion coupling with external laser fields.

The symmetries of the TQRM are controlled by the Rydberg excitation laser. At vanishing detuning the TQRM has a well defined parity symmetry that relies on the total population of the spins and phonon. When the detuning is finite, the parity symmetry disappears. We derive analytically the $\mathcal{G}$-function that defines the eigenspectrum of the TQRM, and show that these analytical results indeed match those of numerical calculations. We then investigate the groundstate properties, focusing on the strong coupling regime. At zero detuning the TQRM reduces to an effective QRM, which differs from the QRM insofar that it describes a collective interaction between the spins and the bosonic mode. In~the superradiant phase the spatial distribution of the phonon splits into two peaks symmetrically, and its state is captured by a classical mixture of two coherent states $\lvert\pm\alpha\rangle$ ($\alpha>0$). At finite detuning the superradiance gradually dominates the groundstate. It~is found that the phonon spatial distribution has a single peak, which is shifted from the origin in the strong coupling regime. A careful examination shows that the phonon state is approximately described by coherent state $\lvert-\alpha\rangle$.
	
This~article is organized as follows. In \Cref{sec:Model&Methods} we present the trapped ion system used as the quantum simulator and derive the TQRM Hamiltonian. In~\cref{sec:Symmetries} we discuss the symmetries present in the model and how the single spin detuning affects the symmetry. The $\mathcal{G}$-function is derived analytically in \cref{sec:spectra}. In~\cref{sec:superradiance} we discuss superradiance and the phase space distribution function of phonons in the groundstate of the TQRM. We~conclude in \cref{sec:conclusions}.

\section{Tripartite quantum Rabi model with a pair of trapped Rydberg ions\label{sec:Model&Methods}}
In this section we provide details on how to realize the TQRM with a Rydberg ion quantum simulator. We consider trapped ions excited to highly excited electronic states with principal quantum number $n\gg 1$. Trapped Rydberg ions exhibit exaggerated properties compared to the groundstate~\cite{Mueller_2008,Schmidt-Kaler_2011}, such as long lifetime~\cite{Saffman_2010}, strong and controllable two-body interactions~\cite{Zhang_2020a}, and strong coupling between vibrational (phonon) and electronic states~\cite{li_electronically_2012,li_parallel_2013,Li_2013, Gambetta_2019, Gambetta_2020, Gambetta_2021}. As~a consequence of the tunable nature of these properties trapped Rydberg ions have been proposed as analogue and digital quantum computers~\cite{Mueller_2008, Li_2013, Gambetta_2019, Zhang_2020a, Gambetta_2020},  platforms for investigating mesoscopic physics~\cite{Lourenco_2022}, exploring topological physics~\cite{Gambetta_2021}, and vibronic coupling~\cite{wilkinson_spectral_2023}. Recently Rydberg excitation of~\ce{^88Sr+} and~\ce{^40Ca+} ions, and entangling gates with trapped Rydberg ions~\cite{Zhang_2020a} have been achieved experimentally~\cite{Feldker_2015, Bachor_2016, [{}] [{, and~the references within.}] Higgins_2018, Mokhberi_2019, [{}] [{, and~the references and reproduced papers within.}] Zhang_2020b, Mokhberi_2020, Andrijauskas_2021}.
\begin{figure}[tb!]
	\includegraphics[width = 1.0\linewidth]{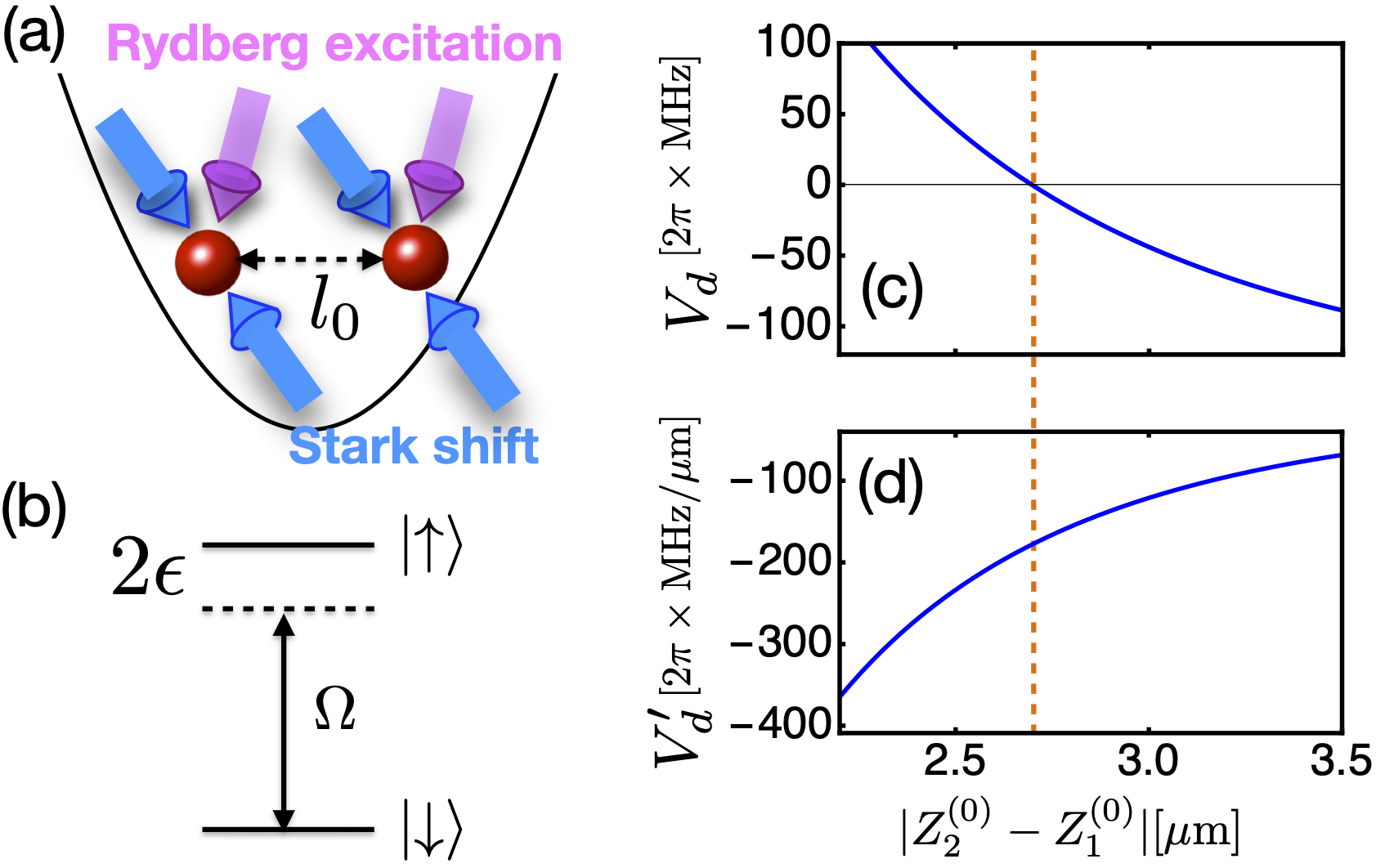}
	\centering
	\caption{(a) Two ions are confined in a linear Paul trap with trap frequency $\nu$. The ions are coherently laser excited to Rydberg states. In their Rydberg states the ions interact through dipolar interaction depending on their separation $l_0=\big|Z_2^{(0)}-Z_1^{(0)}\big|$. A standing wave laser field (blue arrows) induces a site dependent Stark  shift. (b) The Rydberg excitation laser excites the ions from groundstate~$\lvert\downarrow\rangle$ to Rydberg state~$\lvert\uparrow\rangle$ with Rabi frequency~$\Omega$ and detuning~$2\epsilon$. (c) The dipolar interaction. At $l_0= \SI{2.7}{\micro\meter}$ the interaction $V_\mathrm{d}(l_0)=0$ (marked by the dashed perpendicular line). (d) The slope $V_\mathrm{d}^\prime$ of the dipolar interaction.  $V_\mathrm{d}^\prime=-2\pi\times \SI{174.7}{\mega\hertz/\micro\meter}$ when $l_0= \SI{2.7}{\micro\meter}$. The dipolar interaction is induced through using the microwave dressed Rydberg states~\cite{Gambetta_2021}. In this example, we have considered \ce{^88Sr+} with principal quantum number $n=80$, and $\nu = 2\pi\times \SI{2.02}{\mega\hertz}$. See text for details. \label{fig:Schematic}}
\end{figure}

In our setting two ions are trapped in a linear Paul trap, as depicted in \cref{fig:Schematic}(a). Due to the harmonic trapping potential (axial trapping frequency $\nu$) and Coulomb repulsion  the two ions form a Wigner crystal along the trap axis ($z$-axis) at low temperatures, where coordinates of the $j$-th ion are $X_j,Y_j$, and $Z_j$ ($j=1,2$). The ions vibrate around their equilibrium positions $X^{(0)}_j=Y^{(0)}_j=0$ and $Z^{(0)}_2=-Z^{(0)}_1=\big(\mathcal{N}^2e^2/16\pi\epsilon_0M\nu\big)^{1/3}$, where~$\mathcal{N}e$ is the net charge, $M$~is the mass of the ion, and $\epsilon_0$ the permittivity of free space. As we will show later, the electronic states (spin) will couple to the axial vibration, while the radial motions will not be involved. Along the trap axis we obtain a center-of-mass (c.m.)\ and breathing modes whose frequencies are $\omega_\text{c.m.}=\nu$ and $\omega=\sqrt{3}\nu$, respectively~\cite{James_1998}. A Rydberg excitation laser couples the low energy metastable state~${\vert\mathopen\downarrow\rangle} = \lvert 4D_{5/2}\rangle$ and a Rydberg state~${\vert\mathopen\uparrow\rangle}$ (e.g.\ $\lvert nS_{1/2}\rangle$ with $n$ to be the principal quantum number) coherently~\cite{higgins_single_2017}. At the same time an off-resonant standing wave laser induces a state-dependent Stark shift $F_j$~\cite{PhysRevLett.92.207901,deng_effective_2005}. The Hamiltonian of the system reads ($\hbar \equiv 1$),
\begin{equation*}
	H_t = \omega_\text{c.m.}a_\text{c.m.}^\dagger a_\text{c.m.} + \omega a^\dagger a + \sum_{j=1,2}(\Omega \sigma_j^x+ \epsilon\sigma_j^z +F_j) +H_\mathrm{d},
\end{equation*}
where~$a_\text{c.m.}^{\dagger}$ ($a_\text{c.m.}$) and $a^{\dagger}$ ($a$) are the creation (annihilation) operators~of the c.m.\ and breathing mode, respectively. The~operators $\sigma^x_j$, $\sigma^z_j$, and $n_j = \frac{1}{2}\big(1 + \sigma^z_j\big)$ are Pauli and projection operators of the $j$-th ion.  $\Omega$ and~$\epsilon$ are the Rabi frequency and detuning. $F_j$ describes the Stark shift due to the standing wave laser. In Rydberg states the ions interact via a dipole-dipole interaction $H_\mathrm{d}=V_\mathrm{d}(Z_2-Z_1)n_1n_2$ with the distance dependent strength $V_\mathrm{d}(Z_2-Z_1)$~\cite{Li_2013,Zhang_2020a}.

At low temperatures the characteristic length of the c.m. and breathing vibrations  are $l_{\text{c.m.}}=1/\sqrt{2M\nu}$ and $l_\text{b}=1/\sqrt{2M\omega}$, respectively. For typical trap parameters, the length is in the order of a few tens of nanometers, while the distance between the two ions is several micrometers~\cite{James_1998,Zhang_2020a}. This allows us to expand $V_\mathrm{d}(Z_2-Z_1)$ around their equilibrium positions. To the linear order of $Z_j$, we obtain $V_\mathrm{d}(Z_2-Z_1)\approx V_\mathrm{d}(l_0) + V^\prime_\mathrm{d}(l_0)(z_2-z_1)$ where  $V_\text{d}(l_0)$ and  $V_\mathrm{d}^\prime(l_0)$ are the potential and its slope at the equilibrium distance $l_0=\big|Z_2^{(0)}-Z_1^{(0)}\big|$.  $z_j$ is the small deviation from the equilibrium position $Z_j^{(0)}$. By modulating the Rydberg states with external microwave fields we can tune $V_\mathrm{d}(l_0)=0$, i.e.\ the dipolar interaction vanishes at the equilibrium distance as shown in Refs.~\cite{Gambetta_2020,Gambetta_2021}. See \cref{fig:Schematic}(c) and (d) for an example. As a result, the dipolar interaction couples directly to the breathing mode through $V_\mathrm{d}(Z_1-Z_2)\approx G \cdot \big(a^{\dagger}+a\big)$, where $G=l_bV^\prime_\mathrm{d}(l_0)$ gives the coupling strength. Note that the c.m.\ mode decouples with the electronic dynamics. We will exclusively focus on the coupled dynamics between the spin and breathing mode from now on. 

With the above consideration one obtains $H_\mathrm{d} \approx g (\sigma_1^z\sigma_2^z + \sigma_1^z +  \sigma_2^z +1)(a+a^{\dagger})$, with $g=G/4$. Our aim is to achieve a spin-spin-phonon coupled interaction $\propto \sigma_1^z\sigma_2^z(a+a^{\dagger})$. To achieve this, one can turn off the coupling between the individual spin and phonon through spin-dependent Stark shift~\cite{PhysRevLett.92.207901}  with $F_1 = -g(\sigma_1^z+1/2)(a+a^{\dagger})$ and $F_2 = -g(\sigma_2^2+1/2)(a+a^{\dagger})$. The detail of the implementation can be found in the Appendix. The only remaining term is a collective coupling between the two spins and phonon, i.e.~the breathing mode couples to the two spins simultaneously whose strength depends on the slope of the two-body dipolar interaction. This yields the Hamiltonian
\begin{equation}
    H = \omega a^\dagger a + \Omega\big(\sigma_1^x+\sigma_2^x\big) +\epsilon\big(\sigma_1^z+\sigma_2^z\big) +g \big(a^{\dagger}+a\big) \sigma_1^z\sigma_2^z.
    \label{eq:EQRM}
\end{equation}
 \Cref{eq:EQRM} represents the tripartite quantum Rabi model (TQRM) consisting of a monochromatic bosonic (phonon) mode of frequency~$\omega$ and two identical spins. The~unique aspect of the TQRM compared to the QRM and its variants previously studied is that our model is characterized by a tripartite coupling, i.e.~a spin-spin-boson coupling.  The tripartite coupling is different from existing models. In conventional multi-spin QRMs the two-body interactions between spins are typically not present (see Refs.~\cite{Peng_2012, Chilingaryan_2013, Peng_2013, Wang_2014, Peng_2014, Rodriguez-Lara_2014, Duan_2015, Mao_2015, Sun_2020, Yan_2021}), or the spin-spin interactions do not directly couple to the bosonic mode~\cite{Grimaudo_2023, Grimaudo_2023b}. 

\section{Symmetries of the TQRM\label{sec:Symmetries}}
We~begin by examining symmetries in the TQRM and how the single body detuning affects the symmetries. It is convenient to rotate the Hamiltonian~\eqref{eq:EQRM} around the $\sigma_i^y$-axes by $\pi/2$, yielding
\begin{equation}\label{eq:EQRM_rot}
	H_\mathrm{R} = \bosonic a^\dagger a - \qubit\big(\sigma_1^z + \sigma_2^z\big) + \asymmetry\big(\sigma_1^x + \sigma_2^x\big) + \coupling\sigma_1^x\sigma_2^x\big(a^\dagger + a\big) \text{.}
\end{equation}
In the rotated basis the spin up and spin down states will be denoted with $\lvert \Uparrow\rangle$ and $\lvert \Downarrow\rangle$, respectively. After the rotation both the detuning and the tripartite coupling  term of $H_\mathrm{R}$ can induce transitions between spin states.  In the following, we first analyze  symmetries in the TQRM when the Rydberg excitation laser is resonant ($\asymmetry = 0$), where transition between states is solely driven by the tripartite coupling.  Next we consider finite detuning ($\epsilon \neq 0$), where both terms will be taken into account.

\subsection{Resonant TQRM\label{sec:sym_symmetries}}

When $\epsilon=0$, the spin states of the two ions are only coupled through the tripartite interaction, i.e. via $\sigma_1^x\sigma_2^x$ term. The double spin down state $\lvert\mathopen\Downarrow\Downarrow\rangle$ can only couple to $\lvert\mathopen\Uparrow\Uparrow\rangle$ state, i.e. $\sigma_1^x\sigma_2^x\lvert\mathopen\Uparrow\Uparrow\rangle=\lvert\mathopen\Downarrow\Downarrow\rangle$, and vice versa. Here $\lvert s_1s_2\rangle=\lvert s_1\rangle\otimes|s_2\rangle$ with the spin state $|s_j\rangle$ ($s_j = \{\Uparrow,\Downarrow\})$.  A similar transition can be found between states ${\lvert\mathopen\Uparrow\Downarrow\rangle}$ and ${\lvert\mathopen\Downarrow\Uparrow\rangle}$. This shows that the two sectors~$\{{\lvert\mathopen\Uparrow\Uparrow\rangle},{\lvert\mathopen\Downarrow\Downarrow\rangle}\}$ and  $\{{\lvert\mathopen\Uparrow\Downarrow\rangle},{\lvert\mathopen\Downarrow\Uparrow\rangle}\}$ are separable (e.g. block diagonal),  and hence can be treated individually.

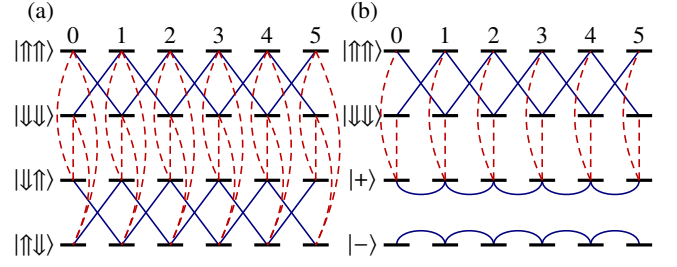
\begin{figure}[tb!]
	\begin{tikzpicture}[scale = 4.3]
		\draw[color = white] (0,0.31) -- (1,0.31);
		\draw[color = darkBlue, semithick] (0.9/6,0.4) -- (1.8/6,0.4+0.2) -- (2.7/6,0.4) -- (3.6/6,0.4+0.2) -- (4.5/6,0.4) -- (5.4/6,0.4 + 0.2);
		\draw[color = darkBlue, semithick] (0.9/6,0.4+0.2) -- (1.8/6,0.4) -- (2.7/6,0.4+0.2) -- (3.6/6,0.4) -- (4.5/6,0.4+0.2) -- (5.4/6,0.4);
		\draw[color = darkBlue, semithick] (0.9/6,0) -- (1.8/6,0+0.2) -- (2.7/6,0) -- (3.6/6,0+0.2) -- (4.5/6,0) -- (5.4/6,0 + 0.2);
		\draw[color = darkBlue, semithick] (0.9/6,0+0.2) -- (1.8/6,0) -- (2.7/6,0+0.2) -- (3.6/6,0) -- (4.5/6,0+0.2) -- (5.4/6,0);
		\node(spin_label_4) at (0.9/6-0.04-0.08 ,0.6){\small$\vert\mathopen{}\Uparrow\Uparrow\rangle$};
		\node(spin_label_3) at (0.9/6-0.04-0.08,0.4){\small$\vert\mathopen{}\Downarrow\Downarrow\rangle$};
		\node(spin_label_1) at (0.9/6-0.04-0.08 ,0){\small$\vert\mathopen{}\Uparrow\Downarrow\rangle$};
		\node(spin_label_2) at (0.9/6-0.04-0.08,0.2){\small$\vert\mathopen{}\Downarrow\Uparrow\rangle$};
		\node(a) at (0.3/6,0.72) {\small(a)};
		\node(0) at (0.9/6,0.65){\small0};
		\node(0) at (1.8/6,0.65){\small1};
		\node(0) at (2.7/6,0.65){\small2};
		\node(0) at (3.6/6,0.65){\small3};
		\node(0) at (4.5/6,0.65){\small4};
		\node(0) at (5.4/6,0.65){\small5};
		\draw[color = white] (0+1,0.31) -- (1+1,0.31);
		\foreach \x in {0.9/6+1,1.8/6+1,2.7/6+1,3.6/6+1,4.5/6+1}{
			\draw[color = darkBlue, semithick] (\x,0) to [out = 90, in = 90] (\x+0.9/6,0);}
		\foreach \x in {0.9/6+1,1.8/6+1,2.7/6+1,3.6/6+1,4.5/6+1}{
			\draw[color = darkBlue, semithick] (\x,0.2) to [out = 270, in = 270] (\x+0.9/6,0.2);}
		\draw[color = darkBlue, semithick] (0.9/6+1,0.4) -- (1.8/6+1,0.6) -- (2.7/6+1,0.4) -- (3.6/6+1,0.6) -- (4.5/6+1,0.4) -- (5.4/6+1,0.6);
		\draw[color = darkBlue, semithick] (0.9/6+1,0.6) -- (1.8/6+1,0.4) -- (2.7/6+1,0.6) -- (3.6/6+1,0.4) -- (4.5/6+1,0.6) -- (5.4/6+1,0.4);
		\foreach \y in {0,0.2,0.4,0.6}{
			\foreach \x in {0.9/6+1,1.8/6+1,2.7/6+1,3.6/6+1,4.5/6+1,5.4/6+1}{
				\draw[very thick] (\x-0.04,\y) -- ((\x+0.04,\y);}}
		\foreach \y in {0,0.2,0.4,0.6}{
			\foreach \x in {0.9/6,1.8/6,2.7/6,3.6/6,4.5/6,5.4/6}{
				\draw[very thick] (\x-0.04,\y) -- ((\x+0.04,\y);}}
		\node(0) at (0.9/6 + 1,0.65){\small0};
		\node(0) at (1.8/6 + 1,0.65){\small1};
		\node(0) at (2.7/6 + 1,0.65){\small2};
		\node(0) at (3.6/6 + 1,0.65){\small3};
		\node(0) at (4.5/6 + 1,0.65){\small4};
		\node(0) at (5.4/6 + 1,0.65){\small5};
		\node(b) at (0.3/6+1,0.72) {\small(b)};
		\node(spin_label_4) at (0.9/6-0.04-0.065 + 1,0.6){\small$\vert\mathopen\Uparrow\Uparrow\rangle$};
		\node(spin_label_3) at (0.9/6-0.04-0.065 + 1,0.4){\small$\vert\mathopen\Downarrow\Downarrow\rangle$};
		\node(spin_label_1) at (0.9/6-0.04-0.065 + 1,0){\small$\vert\mathopen{}-\rangle$};
		\node(spin_label_2) at (0.9/6-0.04-0.065 + 1,0.2){\small$\vert\mathopen{}+\rangle$};
		\foreach \x in {0.9/6,1.8/6,2.7/6,3.6/6,4.5/6,5.4/6}{
			\draw[darkRed, semithick, densely dashed] (\x,0.2) -- (\x,0.4);
			\draw[darkRed, semithick, densely dashed] (\x,0.4) to[in = 65, out = -65] (\x,0);
			\draw[darkRed, semithick, densely dashed] (\x,0.6) to[in = 115, out = -115] (\x,0.2);
			\draw[darkRed, semithick, densely dashed] (\x,0.6) to[in = 115-50, out = -115+50] (\x,0);
		}
		\foreach \x in {0.9/6+1,1.8/6+1,2.7/6+1,3.6/6+1,4.5/6+1,5.4/6+1}{
			\draw[darkRed, semithick, densely dashed] (\x,0.2) to [out = 115,in = -115] (\x,0.6);
		}
		\foreach \x in {0.9/6+1,1.8/6+1,2.7/6+1,3.6/6+1,4.5/6+1,5.4/6+1}{
			\draw[darkRed, semithick, densely dashed] (\x,0.2) to (\x,0.4);
		}
	\end{tikzpicture}
	\centering
	\caption{Coupling between different spin-phonon states. (a) and (b) illustrate the coupling with basis $\{{\vert\mathopen\Uparrow\Uparrow,n\rangle},{\vert\mathopen\Downarrow\Downarrow,n\rangle}, {\vert\mathopen\Uparrow\Downarrow,n\rangle},{\vert\mathopen\Downarrow\Uparrow,n\rangle}\}$, and $\{{\vert\mathopen\Uparrow\Uparrow,n\rangle},{\vert\mathopen\Downarrow\Downarrow,n\rangle}, {\vert+,n\rangle},{\vert-,n\rangle}\}$, respectively. Solid lines show transitions between the basis induced solely by the tripartite coupling. This is the only coupling in the resonant TQRM (i.e. $\epsilon=0$). When $\epsilon\neq 0$ additional transitions are induced by the single body coupling, indicated by dashed lines. \label{fig:Asymmetric_Trajectories}}
\end{figure}

As there are two spin states in sector $\{{\lvert\mathopen\Uparrow\Uparrow\rangle},{\lvert\mathopen\Downarrow\Downarrow\rangle}\}$, we define collective spin operators $S^z = 
{\lvert\Uparrow\Uparrow\rangle \langle\Uparrow\Uparrow\rvert} - {\lvert\Downarrow\Downarrow\rangle\langle\Downarrow\Downarrow\rvert}$ and~$S^x = {\lvert\Uparrow\Uparrow\rangle\langle\Downarrow\Downarrow\rvert} + {\lvert\Downarrow\Downarrow\rangle\langle\Uparrow\Uparrow\rvert}$. 
Then we project Hamiltonian~(\ref{eq:EQRM_rot}) to this subspace, yielding
\begin{equation}
	H_\mathrm{s} = \bosonic a^\dagger a - 2\qubit S^z + \coupling 
	S^x\big(a^\dagger + a\big)\text{.} \nonumber
\end{equation}
Thus TQRM is equivalent to the conventional QRM in this sector. However, due to the collective nature the effective level separation becomes~$4\Omega$, and the phonon mode couples electronic states of the two spins. In this regime $H_\mathrm{s}$ exhibits a parity symmetry with parity operator $\Pi_\mathrm{p} = \exp(i\pi N_\mathrm{E})$~\cite{Braak_2011}, with the total excitation $N_\mathrm{E} = a^{\dagger}a +S^z$. Its eigenvalues are $1$ ($-1$) when $N_\mathrm{E}$ is an even (odd) integer. It can be shown that $\Pi_{\mathrm{p}} H_{\mathrm{s}}\Pi^{\dagger}_{\mathrm{p}}=H_\mathrm{s}$. Such parity, as depicted in~\cref{fig:Asymmetric_Trajectories}(a), requires that when spin states change, phonon state $\lvert n\rangle$  changes to $\lvert n\pm 1\rangle$.

In the spin sector $\{{\lvert\mathopen\Uparrow\Downarrow\rangle},{\vert\mathopen\Downarrow\Uparrow\rangle}\}$, we note that  $-\Omega\big(\sigma_1^z+\sigma_2^z\big)$ is removed from Hamiltonian $H_\text{R}$ as one obtains zero eigenvalue when applying this term on the spin states $\{{\lvert\mathopen\Uparrow\Downarrow\rangle},{\vert\mathopen\Downarrow\Uparrow\rangle}\}$. Hamiltonian $H_\mathrm{R}$ is simplified to be $H_\mathrm{R}^\prime=\omega a^{\dagger}a + g\sigma_1^x\sigma_2^x (a^{\dagger}+a)$. The coupling in this basis is depicted in \cref{fig:Asymmetric_Trajectories}(a) (lower panel, solid lines). By defining superposition basis ${\lvert\pm\rangle}=\frac{1}{\sqrt{2}}({\lvert\Downarrow\Uparrow\rangle}\pm{\lvert\Uparrow\Downarrow\rangle})$, Hamiltonian $H_\mathrm{R}^\prime$ becomes diagonal in this basis, leading to  $H_{\pm} = \bosonic a^\dagger a \pm g\big(a^\dagger + a\big)$, where $+$ ($-$) corresponds to the superposition state ${\lvert+\rangle}$ (${\lvert -\rangle}$). In each superposition state ${\lvert\pm\rangle}$, phonon states~$\lvert n\rangle$ couple to $\lvert n\pm 1\rangle$ without affecting the spin state, as depicted in \cref{fig:Asymmetric_Trajectories}(b). Hamiltonian $H_{\pm}$ describes displaced harmonic oscillators, whose eigenenergies, $E_{\pm}(n)=n\omega-g^2/\omega$, are doubly degenerate at a given $n$~\cite{Irish_2007}.

\subsection{Detuned TQRM\label{sec:aEQRM}}
For finite detuning, i.e.\ $\epsilon\neq 0$, $H_\mathrm{R}$ induces transitions between states $\lvert s_1s_2,n\rangle$ through both the tripartite coupling and $\sigma_j^x$ operators. These couplings are shown with solid and dashed lines in \cref{fig:Asymmetric_Trajectories}(a), respectively. The total spin operator $\mathbf{S}=(\mathbf{s}_1+\mathbf{s}_2)$ is a conserved quantity with $\mathbf{s}_j=\sigma_j^x+\sigma_j^y+\sigma_j^z$. By~defining the triplet manifold $\{{\lvert\mathopen\Uparrow\Uparrow\rangle},{\lvert+\rangle},{\lvert\mathopen\Downarrow\Downarrow\rangle}\}$ and singlet manifold {${\lvert -\rangle}$} one finds that the two manifolds decouple. This can be seen in \cref{fig:Asymmetric_Trajectories}(b). Hamiltonian of the singlet manifold is $H_{-}$, whose eigenenergy $E_-(n)$ is independent of the spin degrees of freedom. 

\section{Energy Spectrum\label{sec:spectra}}
Eigenspectra of many variants of QRMs have been obtained analytically~\cite{Xie_2017}, typically through the Braak $\Gfunc$-function. The~$\Gfunc$-functions are derived either through a transformation into the Segal-Bargmann space of complex analytic functions~\footnote{For introduction to this approach see Ref.~\cite{Xie_2017}; for~a mathematically rigorous definition of these Hilbert spaces and their properties see Refs.~\cite{Bargmann_1961, Segal_1963}.}, or~through the Bogoliubov operator approach (BOA) introduced by~\citet{Chen_2012}. We~will employ the BOA as it generates algebraically simpler $\Gfunc$-functions than the Segal-Bargmann approach~\cite{Chen_2012, Wang_2014, Duan_2015}. When $\epsilon=0$ our model reduces to an effective QRM, whose analytical spectrum is known~\cite{Braak_2011}. In the following we will focus on the spectrum of Hamiltonian (\ref{eq:EQRM}) in the triplet sector for $\asymmetry \ne 0$.

\subsection{Bogoliubov transformation and the Braak $\mathcal{G}$-function}
 \label{sec:Asymm_spec}
 
To diagonalize~$H$ in the spin triplet manifold, we define two sets of Bogoliubov operators $A = a + {\coupling}/{\bosonic}$ and $B = a - {\coupling}/{\omega}$. They are shifted bosonic operators and fulfill the commutation relation of bosonic operators, i.e. $[\xi,\xi^{\dagger}]=1$ and $[\xi,\xi]=[\xi^{\dagger},\xi^{\dagger}]=0$ ($\xi=A,B$). They act on the Fock states~$\vert n_A\rangle=\big(A^\dagger\big)^n\vert0_A\rangle/\sqrt{n!}$ and~$\vert n_B\rangle=\big(B^\dagger\big)^n\vert0_B\rangle/\sqrt{n!}$, respectively. Note that $|0_\xi\rangle$ ($\xi=A,B$) is a displaced  phonon vacuum state $\lvert 0\rangle$ of Hamiltonian (\ref{eq:EQRM}),
\begin{equation}
	\vert0_\xi\rangle = \exp\left(-\frac{1}{2}\left(\frac{\coupling}{\bosonic}\right)^2 \mp \frac{\coupling a^\dagger}{\bosonic}\right)\vert0\rangle\text{,} 
\end{equation}
i.e., $\lvert0_\xi\rangle$ is a coherent state, where the sign is $-$ ($+$) when $\xi=A$ ($\xi=B$). The relation between the coherent states~$\vert0_\xi\rangle$~\cite{Glauber_1963} and~$\vert0\rangle$ is important in a later step of our derivation

Using the Bogoliubov operators $A$ and $A^{\dagger}$, and in the triplet manifold, the Hamiltonian~$H$ can be written as
\begin{equation}\label{eq:asym_A}
	H_{\mathrm{A}} = \begin{bmatrix}
		\mathcal{F}_A + 2\asymmetry & \sqrt{2}\qubit & 0 \\
		\sqrt{2}\qubit & \mathcal{D}_A & \sqrt{2}\qubit \\
		0 & \sqrt{2}\qubit & \mathcal{F}_A - 2\asymmetry
	\end{bmatrix}\text{,}
\end{equation}
where $\mathcal{F}_A = \bosonic A^\dagger A - \coupling^2/\bosonic$ and $\mathcal{D}_A = \bosonic A^\dagger A - 2\coupling\big(A^\dagger + A\big) + 3\coupling^2/\bosonic$. Next we expand the eigenstate~$\vert\Psi_A\rangle$ of $H_A$ with
\begin{equation}
	\vert\Psi_A\rangle = \begin{bmatrix}
		\sum_{n = 0}^\infty\sqrt{n!}c_n\vert n_A\rangle \\
		\sum_{n = 0}^\infty\sqrt{n!}d_n\vert n_A\rangle \\
		\sum_{n = 0}^\infty\sqrt{n!}e_n\vert n_A\rangle
	\end{bmatrix}\nonumber,
\end{equation}
where~$c_n$, $d_n$, and~$e_n$ are the expansion coefficients (proportional to probability amplitudes), respectively. Substituting this expansion into the Schr\"{o}dinger equation $H_A\big|\Psi_A\big\rangle = E\big|\Psi_A\big\rangle$ we obtain
\begin{widetext}
	\begin{flalign}
		&\sum_{n = 0}^\infty\left(\bosonic n - \frac{\coupling^2}{\bosonic} + 2\asymmetry\right)\sqrt{n!}c_n\vert n_A\rangle + \sqrt{2}\qubit\sum_{n = 0}^\infty\sqrt{n!}d_n\vert n_A\rangle = E\sum_{n = 0}^\infty\sqrt{n!}c_n\vert n_A\rangle\text{,} \nonumber\\
		&\sum_{n = 0}^\infty d_n\left[\sqrt{n!}\left(n\bosonic + \frac{3\coupling^2}{\omega}\right)\vert n_A\rangle - 2\coupling\Big(\sqrt{n^2}\sqrt{(n - 1)!}\vert(n - 1)_A\rangle + \sqrt{(n + 1)!}\vert(n + 1)_A\rangle\Big)\right] \nonumber\\
		& \hphantom{=} + \sqrt{2}\qubit\sum_{n = 0}^\infty\sqrt{n!}(c_n + e_n)\vert n_A\rangle = E\sum_{n = 0}^\infty\sqrt{n!}d_n\vert n_A \rangle, \nonumber\\
		&\sum_{n = 0}^\infty\left(\bosonic n - \frac{\coupling^2}{\bosonic} - 2\asymmetry\right)\sqrt{n!}e_n\vert n_A\rangle + \sqrt{2}\qubit\sum_{n = 0}^\infty\sqrt{n!}d_n\vert n_A\rangle = E\sum_{n = 0}^\infty\sqrt{n!}e_n\vert n_A\rangle\text{,}\nonumber
	\end{flalign}
		\end{widetext}
	for eigenenergy $E$. Multiplying by the state~$\langle n_A\vert$ from the left hand side, we derive coupled equations of the coefficients,
	\begin{equation}
		\left(\bosonic m - \frac{\coupling^2}{\omega} + 2\asymmetry\right)c_m + \sqrt{2}\qubit d_m = Ec_m\text{,} \label{eq:c_coeff} 
		\end{equation}
		\begin{eqnarray}
		\left(\bosonic m - \frac{3\coupling^2}{\omega}\right)d_m - 2g[(m + 1)d_{m + 1} &-& d_{m - 1}]  \nonumber \\
		+ \sqrt{2}\qubit(c_m +  e_m) &=& Ed_m\text{,} \label{eq:d_coeff} 
		\end{eqnarray}
		\begin{eqnarray}
		\left(\bosonic m - \frac{\coupling^2}{\omega} - 2\asymmetry\right)e_m + \sqrt{2}\qubit d_m = Ee_m\label{eq:e_coeff}\text{.}
	\end{eqnarray}
    From Eq.~(\ref{eq:c_coeff}) and (\ref{eq:e_coeff}) one finds expressions of $c_m$ and $e_m$ with respect to the common coefficient~$d_m$. Substituting these into Eq.~(\ref{eq:d_coeff}) yields a recursion relation for $d_m$,
	\begin{equation*}\label{eq:d_asym}
		md_m = \mathcal{C}_{m - 1}d_{m - 1} - d_{m - 2}\text{,}
	\end{equation*}
	where the coefficient~$\mathcal{C}_m$ is defined as
	\begin{equation}
		\mathcal{C}_m^{(A)} = \frac{1}{2\coupling}\left[\bosonic m + \frac{3\coupling^2}{\bosonic} - E + 2\qubit^2\left(\frac{1}{D + 2\asymmetry} + \frac{1}{D - 2\asymmetry}\right)\right]\text{,} \nonumber
	\end{equation}
with $D = E - \bosonic m + \coupling^2/\bosonic$. Using the initial coefficients $d_0 = 1$ and $d_1 = \mathcal{C}_0$, the~coefficients~$d_m$ can be evaluated iteratively when parameters $\{E,\bosonic,\asymmetry,\qubit,\coupling\}$ are specified.

In a similar way we rewrite the Hamiltonian~$H$ using operators~$B$ and $B^{\dagger}$,
\begin{equation}\label{eq:asym_B}
	H_{B} = \begin{bmatrix}
		\mathcal{F}_B + 2\asymmetry & \sqrt{2}\qubit & 0 \\
		\sqrt{2}\qubit & \mathcal{D}_B & \sqrt{2}\qubit \\
		0 & \sqrt{2}\qubit & \mathcal{F}_B - 2\asymmetry
	\end{bmatrix}\text{,}
\end{equation}
where $\mathcal{F}_B = \bosonic B^\dagger B + 2\coupling\big(B^\dagger + B\big) + 3\coupling^2/\bosonic$ and $\mathcal{D}_B =\bosonic B^\dagger B - \coupling^2/\bosonic$.  The corresponding eigenstate~$\vert\Psi_B\rangle$ is expressed as
\begin{equation}
	\vert\Psi_B\rangle = \begin{bmatrix}
			\sum_{n = 0}^\infty(-1)^n\sqrt{n!}c_n^\prime\vert n_B\rangle \\
			\sum_{n = 0}^\infty(-1)^n\sqrt{n!}d_n^\prime\vert n_B\rangle \\
			\sum_{n = 0}^\infty(-1)^n\sqrt{n!}e_n^\prime\vert n_B\rangle
	\end{bmatrix}. \nonumber
\end{equation}
where the coefficients~$c_n^\prime$, $d_n^\prime$, and $e_n^\prime$ satisfy the recursion relations
\begin{align}
	mc_m^\prime & = \frac{\qubit}{\sqrt{2}\coupling}d_{m - 1}^\prime + \mathcal{C}^+_{m - 1}c_{m - 1}^\prime - c_{m - 2}^\prime\text{,} \label{eq:c_p_coeff} \\
	me_m^\prime & = \frac{\qubit}{\sqrt{2}\coupling}d_{m - 1}^\prime + \mathcal{C}_{m - 1}^-e_{m - 1}^\prime - e_{m - 2}^\prime\text{,} \label{eq:d_p_coeff} \\
	md_m^\prime & = \frac{\sqrt{2}\qubit}{E + \coupling^2/\bosonic -\bosonic m}(c_m^\prime + e_m^\prime)\label{eq:e_p_coeff}\text{,}
\end{align}
with $\mathcal{C}^\pm_m = \left(\bosonic m + 3\coupling^2/{\bosonic} \pm 2\asymmetry - E\right)/{2\coupling}$. 
 
In~the standard BOA approach the arbitrary expansion coefficients in both wavefunctions $\lvert \Psi_A\rangle$ and $\lvert\Psi_B\rangle$ are all defined through recurrence relation of a single coefficient~\cite{Duan_2015, Duan_2015b, Braak_2011, Duan_2022, Chen_2012}. To obtain a recurrence relation from \cref{eq:c_p_coeff,eq:d_p_coeff,eq:e_p_coeff} we first derive initial coefficients $c_0'$ and $e_0'$ by  following the approach presented in Ref.~\cite{Wang_2014}.  We left-multiply wavefunction~$\vert\Psi_A\rangle$ by the bra-state~$\langle0_B\vert$, which yields relations between $c_0'$ ($e_0'$) and $c_n$ ($e_n$),
\begin{subequations}
	\begin{eqnarray}
		c^\prime_0 & = & \ex{-2\coupling^2/\omega}\sum_{n = 0}^\infty c_n\left(\frac{2\coupling}{\bosonic}\right)^n\text{,} \nonumber\\
		e_0^\prime & = & \ex{-2\coupling^2/\omega}\sum_{n = 0}^\infty e_n\left(\frac{2\coupling}{\bosonic}\right)^n\text{.} \nonumber
	\end{eqnarray}
\end{subequations}
When deriving the above expressions, we have used the relation $\langle0_B\vert n_A\rangle = \frac{1}{\sqrt{n!}}\big(\frac{2\coupling}{\bosonic}\big)^n\ex{-2\coupling^2/\bosonic^2}$. 
With these initial coefficients $c_0'$ and $e_0'$ the first iteration is obtained,
\begin{equation}
	c_1^\prime = \frac{\qubit}{\sqrt{2}\coupling}d_0^\prime + \mathcal{C}_0^+c_0^\prime \qquad\text{and}\qquad e_1^\prime = \frac{\qubit}{\sqrt{2}\coupling}d_0^\prime + \mathcal{C}_0^-e_0^\prime, \nonumber
\end{equation}
which allow us to solve all expansion coefficients iteratively through \cref{eq:c_p_coeff,eq:d_p_coeff,eq:e_p_coeff}.

If~$E$ is a nondegenerate eigenenergy of~$H$ the states $\vert\Psi_A\rangle$ and~$\vert\Psi_B\rangle$ must differ only by a complex coefficient~$K$, i.e. $\vert\Psi_A\rangle= K \vert\Psi_B\rangle$. Left-multiplying by the vacuum bra-state~$\langle0\vert$ on both sides, we obtain 
\begin{subequations}
	\label{eq:assym_equality}
	\begin{eqnarray} 
		\sum^\infty_{n = 0}c_n\ex{-\coupling^2/2\bosonic^2}\left(\frac{\coupling}{\bosonic}\right)^n & = & K\sum^\infty_{n = 0}c_n^\prime\ex{-\coupling^2/2\bosonic^2}\left(\frac{\coupling}{\bosonic}\right)^n\text{,}\qquad \\
		\sum^\infty_{n = 0}e_n\ex{-\coupling^2/2\bosonic^2}\left(\frac{\coupling}{\bosonic}\right)^n & = & K\sum^\infty_{n = 0}e_n^\prime\ex{-\coupling^2/2\bosonic^2}\left(\frac{\coupling}{\bosonic}\right)^n\text{,}
	\end{eqnarray}
\end{subequations}
where we have used relation $\sqrt{n!}\langle0\vert n_A\rangle = (-1)^n\sqrt{n!}\langle0\vert n_B\rangle = \ex{-\coupling^2/2\bosonic^2}\left({\coupling}/{\bosonic}\right)^n$~\cite{Chen_2012}. Cross-multiplying to eliminate the arbitrary constant $K$ an analytical expression for the Braak $\Gfunc$-function is found,
\begin{flalign}\label{eq:G_func_as}
	\Gfunc(E) & = \sum^\infty_{n = 0}c_n\left(\frac{\coupling}{\bosonic}\right)^n\sum^\infty_{n = 0}e_n^\prime\left(\frac{\coupling}{\bosonic}\right)^n \nonumber \\
	& \hphantom{=} - \sum^\infty_{n = 0}c_n^\prime\left(\frac{\coupling}{\bosonic}\right)^n\sum^\infty_{n = 0}e_n\left(\frac{\coupling}{\bosonic}\right)^n\text{.}
\end{flalign}
The roots of the $\mathcal{G}$-function are used to evaluate eigenenergies of Hamiltonian~(\ref{eq:EQRM}).

\begin{figure}[tb!]
	\includegraphics[width = 1.0\linewidth]{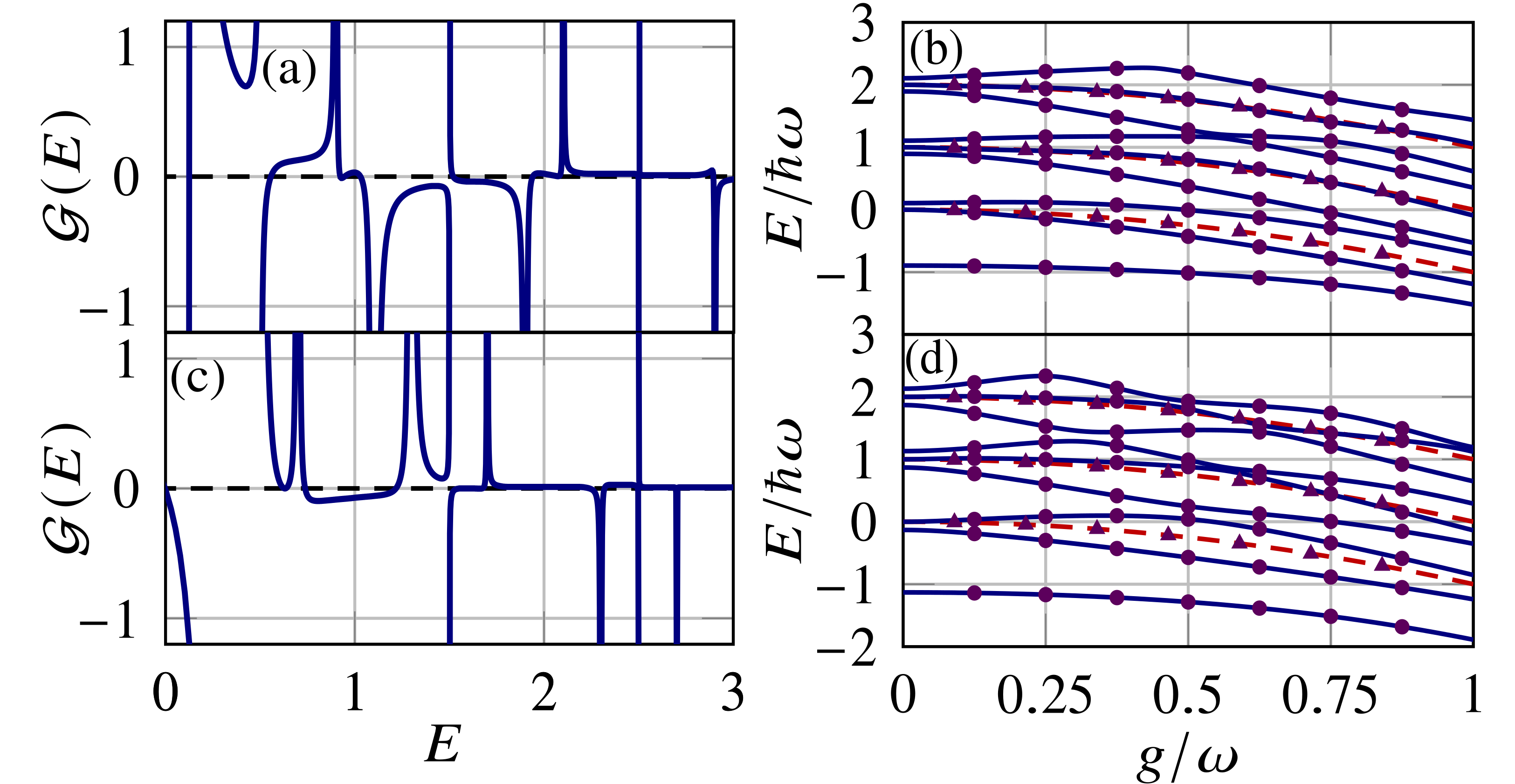}
	\centering
	\caption{(a) Braak $\Gfunc(E)$ function and (b) energy spectra of the TQRM for $\asymmetry/\bosonic = 0.2$.  (c) Function $\Gfunc(E)$ and  (d) eigenspectra  for $\asymmetry/\bosonic = 0.4$. In (b) and (d) the~triplet and singlet eigenenergies are denoted with solid blue lines and dashed red lines, respectively. Purple dots and triangles denote analytical results from the $\Gfunc$-function and $E_\pm(n)$, respectively. In (a) and (c), we~have set $\qubit/\bosonic = 0.4$ and $\coupling/\bosonic = 0.5$.\label{fig:Asymmetric_G_function}}
\end{figure}

Examples of the function $\mathcal{G}(E)$ are plotted in \cref{fig:Asymmetric_G_function}(a) and (c). Unlike the TQQRM~\cite{Wang_2014, Duan_2015}, there are no regular pole structures in our model. This results exclusively from the detuning term that breaks the symmetry in the QRM~\cite{Shi_2022,Braak_2011}. Roots of the $\mathcal{G}(E)$ function are evaluated numerically. In \cref{fig:Asymmetric_G_function}(b) eigenvalues of $H$ as a function $g/\omega$ are shown. The eigenenergy obtained from the $\mathcal{G}(E)$ function agrees with the numerical diagonalization of the Hamiltonian.  Anticrossings of the spectra are found at finite $g$. However the groundstate energy clearly separates from the first excited state. The energy gap between the ground and first excited state remains finite when $g/\omega > 1$. Increasing $\epsilon$ the gap increases too, as can be seen in \cref{fig:Asymmetric_G_function}(b) and (d). 

\section{Subradiant and superradiant phases in the groundstate\label{sec:superradiance}}
In this section we investigate properties of the groundstate of the TQRM. Through mean-field calculations and numerical diagonalization of the full Hamiltonian we will show that a subradiant to superradiant transition can be induced by increasing $g$ in the resonant TQRM. Phase space densities of the phonon state are symmetric but highly delocalized in the superradiant regime. The phonon state can be described by a classical mixture of coherent states ${|\mathopen\pm\alpha\rangle}$. In the detuned TQRM we will show that the phonon state is a displaced oscillator. The subradiance disappears, as the mean phonon number is non-zero as long as $g>0$ in the detuned TQRM. 

\subsection{Delocalized superradiant phase of the resonant TQRM\label{sec:superradiance_symmetric}}
\begin{figure}[tb!]
	\includegraphics[width = 0.9\linewidth]{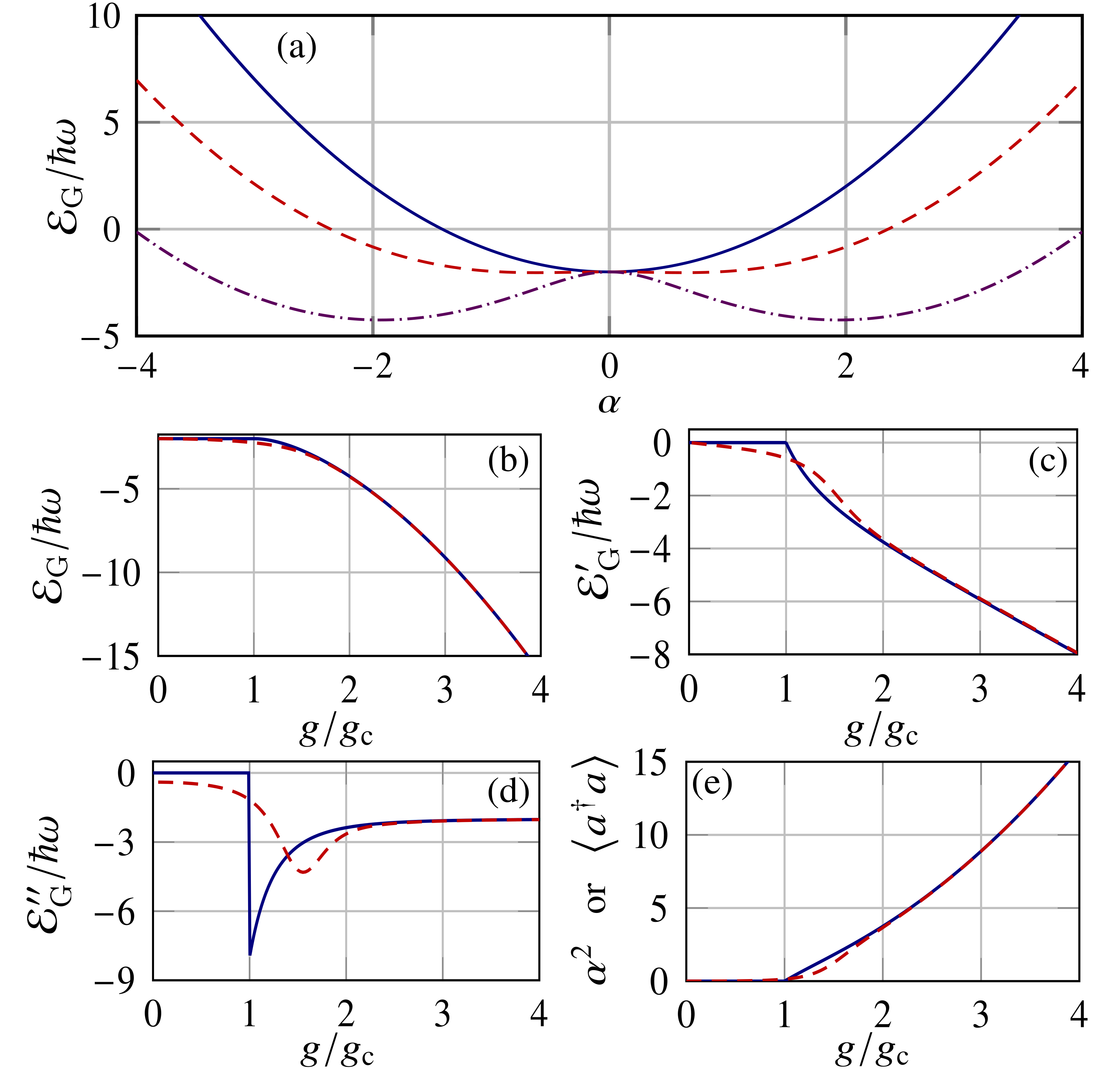}
	\centering
	\caption{(a) Mean-field  energy functional~$\mathcal{E}_\mathrm{G}$ for $g/g_\mathrm{c} = 0.0$ (solid blue), $1.1$ (dashed red), and~$2.0$ (dash-dotted purple). The groundstate energy is obtained at the minima of the curves.  (b) The~groundstate energy, (c) its~first and (d) second derivatives, as well as~(e) the population of the bosonic mode~$\big\langle a^\dagger a\big\rangle = \alpha^2$ with respect to the coupling. Solid (dashed) lines represent the mean-field (quantum) results. In the following calculations, we scale the Hamiltonian with respect to $\bosonic$ and set $\Omega=\bosonic$.\label{fig:nature_of_transition}}
\end{figure}

The groundstate exhibits a subradiant to superradiant phase transition when $g$ is larger than a critical value. We first determine the critical value within the mean-field approach. This is done by~replacing operators $a$ ($a^{\dagger}$) with their mean value $\alpha$ ($\alpha^*$) in Hamiltonian \cref{eq:EQRM}. By assuming that $\alpha$ is real we diagonalize the Hamiltonian, and~the lowest energy $\mathcal{E}_{\text{G}} =  \bosonic\alpha^2 - 2\sqrt{\coupling^2\alpha^2 + \Omega^2}$ is obtained as a function of $\alpha$. In the groundstate $\partial \mathcal{E}_\mathrm{G}/\partial\alpha = 0$, which allows the definition of a critical coupling $g_\mathrm{c} = \sqrt{\bosonic\qubit}$. When $g \le g_\mathrm{c}$ we find $\alpha=0$. The groundstate is subradiant with zero phonon occupation. When $\coupling > \coupling_\mathrm{c}$, two non-zero solutions, $\alpha=\pm\alpha_0$ with $\alpha_0=\sqrt{{\coupling^2}/{\bosonic^2} - {\qubit^2}/{\coupling^2}}$, are obtained indicating superradiance with a finite phonon occupation. As shown in \cref{fig:nature_of_transition}(a), $\mathcal{E}_\mathrm{G}$ is an even function of~$\alpha$. Two  minima are found when $g>g_c$, consistent with the above analysis. 

At the mean-field level this superradiant transition is characteristic of a spontaneous symmetry breaking occurring related to a second order transition~\cite{Hwang_2015, Chen_2020}. Here the energy (Fig.~\ref{fig:nature_of_transition}(b)) and its first derivative (Fig.~\ref{fig:nature_of_transition}(c)) are continuous when varying $g$. However the second derivative of the groundstate energy shows a discontinuity at the critical coupling, as seen in Fig.~\ref{fig:nature_of_transition}(d).  Both the mean-field and the exact diagonalization calculation agree well in the subradiant and superradiant phases. Around the critical point $g=g_\text{c}$, the full quantum model shows a crossover, which is different from the sharp transition in the mean-field calculation.   
   
\begin{figure}[tb!]
	\includegraphics[width = 0.98\linewidth]{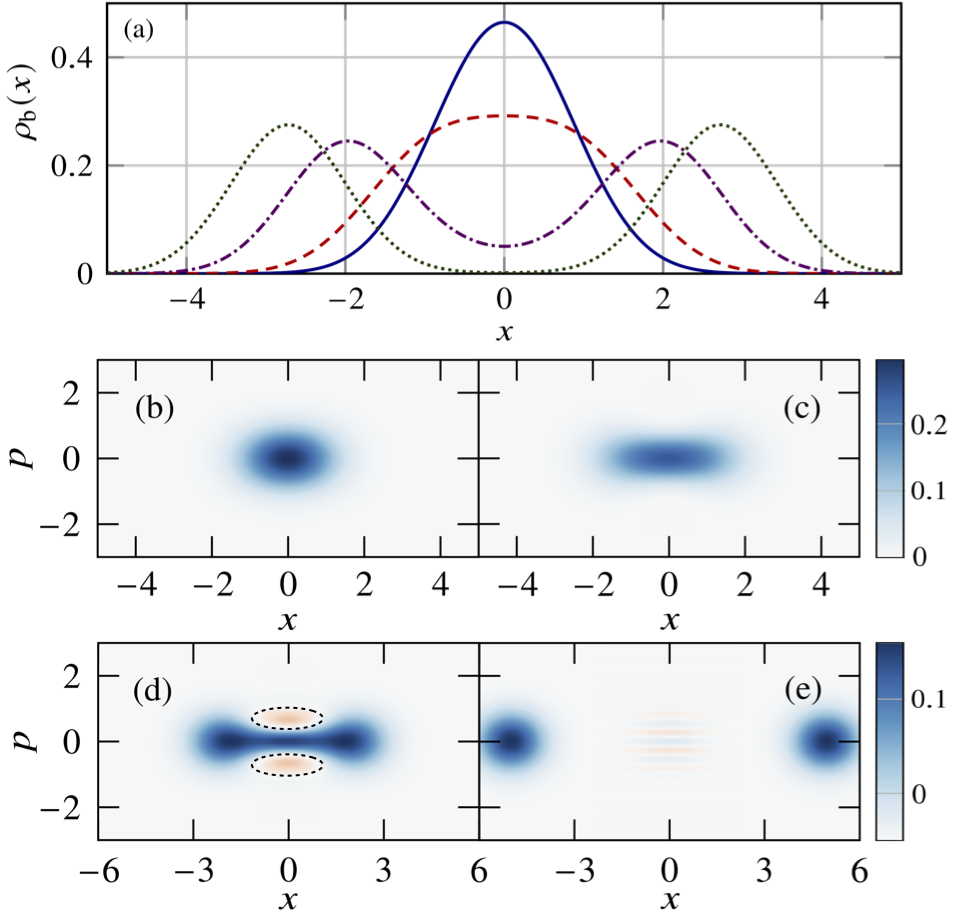}
	\centering
	\caption{(a)  Density distribution~$\rho_\mathrm{b}(x)$ for couplings $g/g_\mathrm{c} = 0.8$ (solid), $g/g_\mathrm{c} = 1.2$ (dashed), $g/g_\mathrm{c} = 1.6$ (dash-dotted), and~$g/g_\mathrm{c} = 2.0$ (dotted). Delocalization of the spatial density is found in the superradiant phase. The Wigner distribution $W(x,p)$ for (b) $g/g_\mathrm{c} = 0.8$, (c) $g/g_\mathrm{c} = 1.2$, (d) $g/g_\mathrm{c} = 1.6$, and~(e) $g/g_\mathrm{c} = 3.5$.  The Wigner distribution is stretched along the $x$-axis when increasing $g$. In the superradiant regime, negative values are found, see regions marked by the ellipses in (d). In the strong coupling regime, the negative region is negligible. \label{fig:wavefunctions_symmetric}}
\end{figure}

We now investigate the phonon states in the subradiant and superradiant phase. We are particularly interested in the phonon properties in the position representation. The position and momentum are given through phonon operators $x=l_\mathrm{b}(a^{\dagger}+a)$ and $p= \frac{\mathrm{i}}{l_\mathrm{b}}(a^{\dagger}-a)$, respectively. We first numerically evaluate the reduced density matrix of the phonon $\rho_{\text{b}}=\tr[\rho_\mathrm{t}]$, by tracing the spin degree of freedom from the total density matrix $\rho_\mathrm{t} = |\psi_\mathrm{G}\rangle\langle\psi_\mathrm{G}|$. This allows us to obtain spatial density distribution $\rho_\text{b}(x)=\langle x|\rho_\text{b}|x\rangle$ in the position representation. As shown in Figs.~\ref{fig:wavefunctions_symmetric}(a) the spatial density has a Gaussian profile centered at $x = 0$ in the subradiant phase. When~$g \gtrsim g_\mathrm{c}$ the density deviates from the Gaussian. It~first becomes non-Gaussian around the critical value, and then splits into two separated Gaussian shapes.

The~Wigner quasiprobability distribution $W(x,p)$~\cite{Wigner_1932}, on the other hand, exhibits distinctive features in both phases.  As shown in Figs.~\ref{fig:wavefunctions_symmetric}(b) and (c) the Wigner distribution is stretched along the $x$-axis when $g\sim g_{\mathrm{c}}$. Two separate peaks are observed when $g>g_{\mathrm{c}}$, depicted in \cref{fig:wavefunctions_symmetric}(d)-(e). These peaks are centered around $ x\approx \pm2l_\text{b}\alpha_0$, similar to the spatial density shown in \cref{fig:wavefunctions_symmetric}(a). The Wigner distribution is stretched horizontally (i.e.\ along the $x$ axis)  in the superradiant state. In certain regions, the Wigner function becomes negative, indicating the phonon cannot be described by a Gaussian state. 

\begin{figure}[tb!]
	\includegraphics[width = 1.0\linewidth]{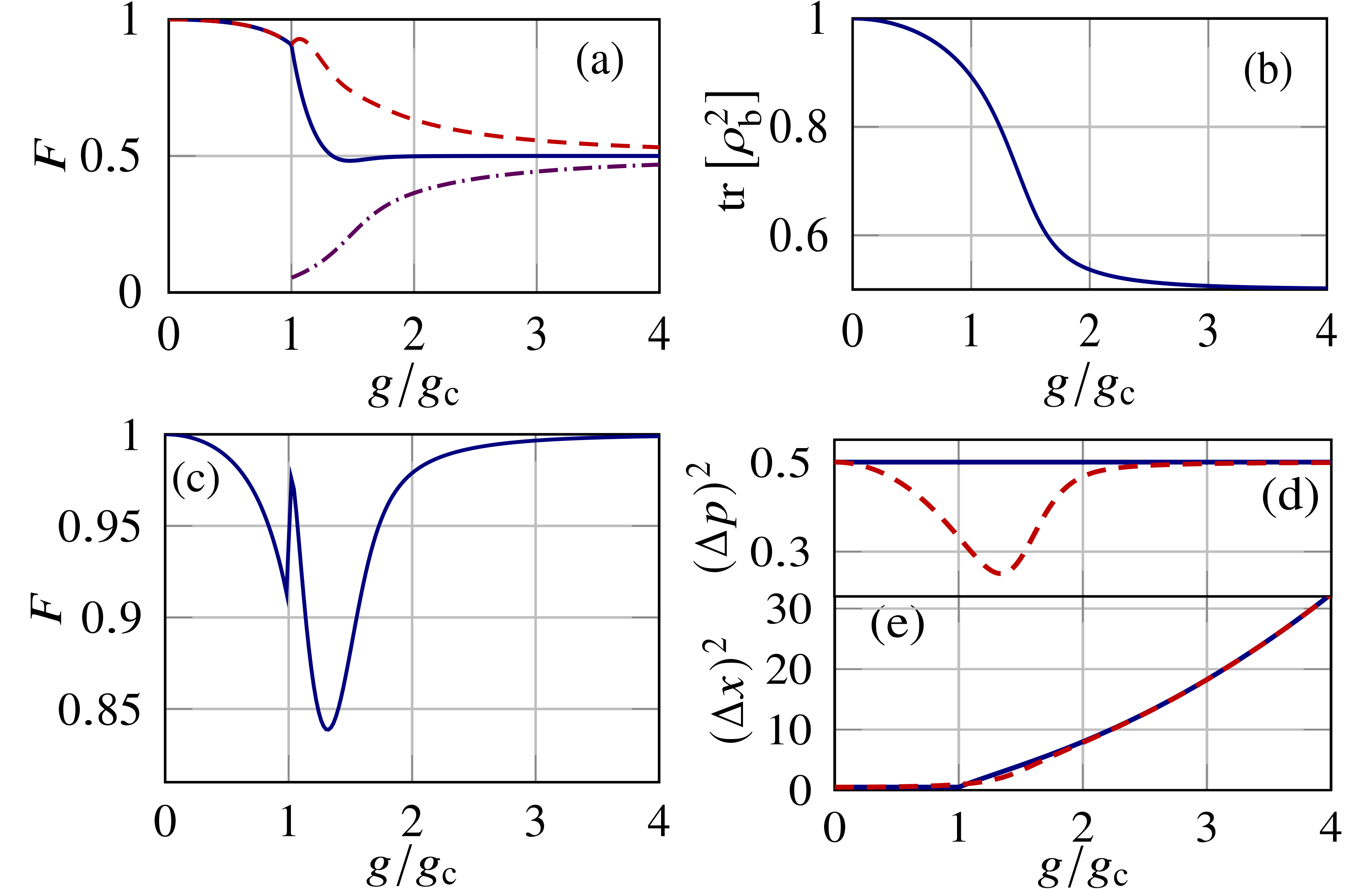}
	\centering
    \caption{(a) The Uhlmann-Jozsa fidelity~$F$ and (b) purity of the  reduced density matrix $\rho_\text{b}$. In~Panel~(a) fidelities~$F$ of $\rho_\text{b}$ and the pure states~${\vert\mathopen\pm\alpha_0\rangle}$ (solid blue), $\vert C_+\rangle$ (dashed red), and~$\vert C_-\rangle$ (dash-dotted purple) are considered for $g>g_\text{c}$. When $g<g_{\text{c}}$, the projection to the vacuum state is examined. In~Panel~(c) the fidelity between $\rho_\text{b}$ and $\rho_{\text{c}}$ is shown. Panels (d) and (e) show the quadrature variances obtained by the mean-field (solid red) and full numerical calculations (dashed blue). \label{fig:fidelities_symmetric}}
\end{figure}

To~understand the nature of the phonon state, we project $\rho_\text{b}$ to well defined reference states characterized by density matrix $\rho_{\text{r}}$. Their overlap is quantified by the Uhlmann-Jozsa fidelity~$F=\left(\text{Tr} \sqrt{\sqrt{\rho_\mathrm{r}}\rho_{\text{b}}\sqrt{\rho_{\text{r}}}}\right)^2$ between density matrix $\rho_{\text{b}}$ and $\rho_{\text{r}}$~\cite{uhlmannTransitionProbabilityState1976a,mendoncaAlternativeFidelityMeasure2008}. When $g<g_\mathrm{c}$, the fidelity is high when $\rho_\mathrm{r}=\lvert 0\rangle\langle 0\rvert$, as shown in \cref{fig:fidelities_symmetric}(a). This is a direct manifestation of the subradiance in this regime. When $g\sim g_\mathrm{c}$, however, the fidelity becomes low in general when projecting $\rho_{\text{b}}$ to ${|\mathopen\pm \alpha_0\rangle}$ or their superpositions $|C_{\pm}\rangle=A_{\pm}\left(\lvert+\alpha_0\rangle \pm\lvert-\alpha_0\rangle\right)$, where $A_{\pm} = 1/\sqrt{2(1 \pm \ex{-2\alpha_0^2})}$ is the normalization constant.  This is consistent with our numerical calculation shown in Fig.~\ref{fig:wavefunctions_symmetric}(c), where the Wigner function exhibits negative values. In this regime, the phonon state is a non-Gaussian state, and hence can not be accurately described by coherent states.

On the other hand, the purity of the phonon density matrix decreases rapidly with increasing $g$ when $g\gg g_\text{c}$, as shown in \cref{fig:fidelities_symmetric}(b).  It turns out that the $\rho_\text{b}$ becomes a classical mixture $\rho_\mathrm{c} = \frac{1}{2}(\lvert+\alpha_0\rangle\langle+\alpha_0\rvert + \lvert-\alpha_0\rangle\langle-\alpha_0\rvert)$.  When $g\gg g_\text{c}$, the corresponding fidelity $F$ approaches unity, as shown in \cref{fig:fidelities_symmetric}(c). Using $\rho_\mathrm{c}$, we can evaluate  the uncertainty of $x$ and $p$ analytically, $(\Delta x)^2=(4\alpha^2+1)/2\omega^2$ and $(\Delta p)^2 = \omega^2[1-(\alpha-\alpha^*)^2]/2$. As $\alpha$ is a real number in the mean-field calculation, this leads to $(\Delta p)^2 = \omega^2/2$. The uncertainty $(\Delta p)^2$ is therefore same with that of the coherent state $|{\pm}\alpha_0\rangle$. In other words the phonon state is strongly stretched along the $x$-axis, while the uncertainty along $p$-axis remains a constant. Numerical diagonalization  of the full Hamiltonian shows that~$(\Delta p)^2$ only deviates from~$\omega^2/2$ around $g \sim g_\mathrm{c}$, as can be seen in \cref{fig:fidelities_symmetric}(d).

\subsection{Displaced superradiant phase in the detuned TQRM}
\begin{figure}[tb!]
	\includegraphics[width = 1.0\linewidth]{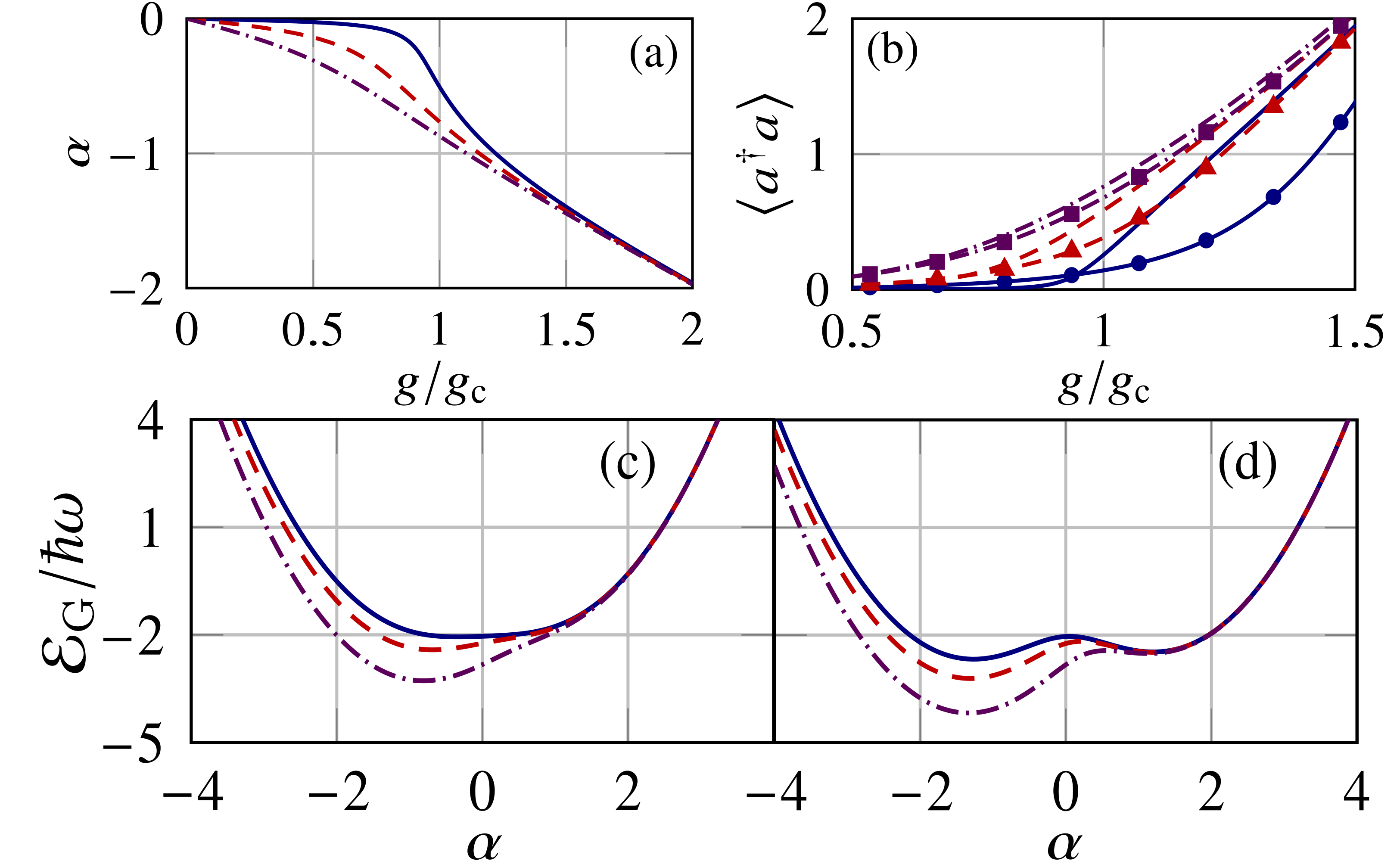}
	\centering
	\caption{(a)~Mean-field~$\alpha$ and (b) phonon population $\langle a^{\dagger}a\rangle$ ($|\alpha|^2$) of the resonant TQRM. In~Panel~(b) the quantum  and mean-field results are denoted by lines and lines with symbols, correspondingly. Energy functional $\mathcal{E}_{\text{G}}$ for (c) $g/g_\mathrm{c} = 0.95$ and (d) $g/g_\mathrm{c} = 1.4$ are shown. In both the subradiant and superradiant regime, minima of the energy functional locate at negative $\alpha$.  We have considered $\asymmetry>0$ in the calculation, where $\asymmetry/\qubit = 0.2$ (solid blue), $\asymmetry/\qubit = 0.5$ (dashed red), and $\asymmetry/\qubit = 1.0$ (dash-dotted purple) in all the panels.\label{fig:Destruciton_of_Spontaneous_symmetry}}
\end{figure}

We first investigate the groundstate of the detuned TQRM with the mean-field approach. In \cref{fig:Destruciton_of_Spontaneous_symmetry}(a) the parameter $\alpha$ as a function of $g$ is shown. When increasing $g>0$ and restricting $\epsilon > 0$ we find $\alpha$ is always negative and its magnitude $|\alpha|$ increases monotonically. This is different from the resonant TQRM where both positive and negative branch of $\alpha$ are found. Another important difference is that the sharp change of $\alpha$ at the subradiant-superradiant transition disappears when $\epsilon > 0$. Instead a smooth crossover emerges around $g \sim g_\mathrm{c}$. We then calculate  the mean phonon population, shown in \cref{fig:Destruciton_of_Spontaneous_symmetry}(b). It is found that the crossover at $g\sim g_\mathrm{c}$ persists. The numerical diagonalization and mean-field results deviate apparently when $\epsilon$ is small. This trend changes when $\epsilon$ is large, where the mean-field result agrees with the numerical calculation and are largely independent of $\epsilon$. Another feature is that when $\epsilon$ is large the phonon number is non zero as long as $g> 0$, indicating that subradiance does not exist any more. In other words the subradiant to superradiant transition is removed by finite $\epsilon$. 

The negativity of $\alpha$ can be understood through the mean-field analysis. When $\epsilon \gg g$ and $\epsilon\gg \Omega$ the groundstate energy functional is well approximated by 
\begin{equation}\label{eq:ground_limit}
	\mathcal{E}_\mathrm{G}\approx \bosonic\alpha^2 + 2\coupling\alpha - 2\asymmetry\text{.} \nonumber
\end{equation}
When plotting $\mathcal{E}_{\text{G}}$ as a function of $\alpha$ [see  Figs.~\ref{fig:Destruciton_of_Spontaneous_symmetry}(c)-(d)] it is apparent that  the minima locate at $\alpha<0$. The value of $\alpha$ can be determined through solving $\partial\mathcal{E}_\mathrm{G}/\partial\alpha=0$. One finds groundstate energy ${E}_\text{G}\approx -g^2/\omega -2\epsilon $ when $\alpha_1 = -\coupling/\bosonic$. This explains qualitatively why $\alpha$ becomes negative in the groundstate. 

\begin{figure}[tb!]
	\includegraphics[width = 1.0\linewidth]{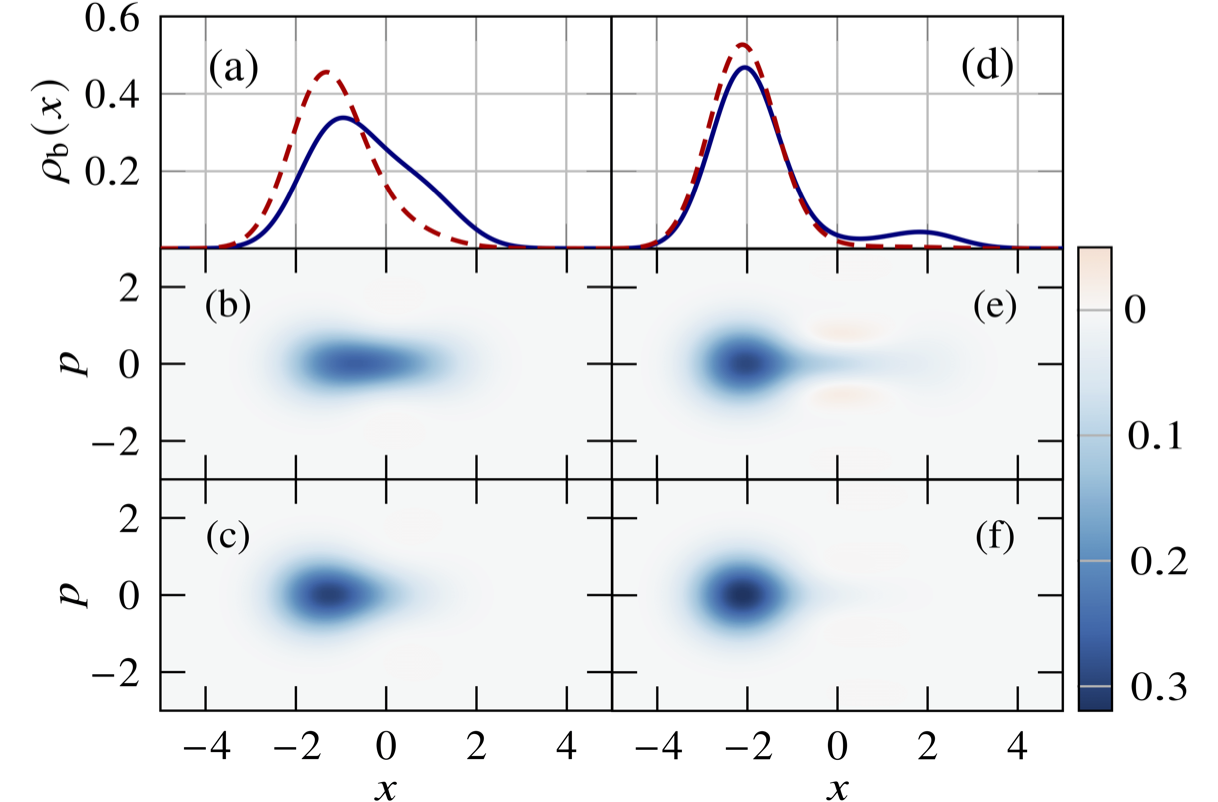}
	\centering
	\caption{Groundstate density and Wigner distribution of the detuned TQRM. When $g/g_\text{c}=1.2$, densities for $\epsilon/\Omega=0.2$ (solid) and $0.4$ (dashed) are shown in panel (a). We show corresponding Wigner distribution~$W(x,p)$ in Panels (b) and (c). In (d), $g/g_\text{c}=1.6$ and we show the corresponding distributions in (e) and (f). The density is localized when $\epsilon/\Omega=0.1$ (solid) and $\epsilon/\Omega=0.2$ (dashed). The corresponding Wigner distribution is shown in panel (e) and (f). \label{fig:Asymmetric_Wigner}}
\end{figure}

The c.m.\ of the phonon density is shifted from $x=0$ (when $\epsilon=0$) towards $x<0$ (when $\epsilon \neq 0$). Our numerical simulations show that the center of the spatial density [\cref{fig:Asymmetric_Wigner}(a)] and Wigner distribution [Figs.~\ref{fig:Asymmetric_Wigner}(b)-(c)] are around $x\approx l_\text{b}\alpha_1$. In addition, the center of the density shifts even further when we increase coupling $|g|$ due to $x \propto -g$. This trend can be seen in Figs.~\ref{fig:Asymmetric_Wigner}(d)-(f).  

\begin{figure}[tb!]
	\includegraphics[width = 1.0\linewidth]{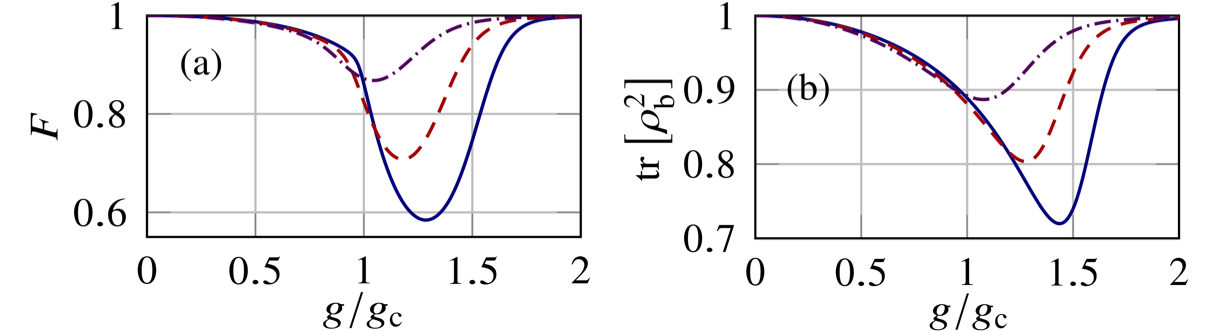}
	\centering
	\caption{(a)~Uhlmann-Jozsa fidelity between $\rho_\text{b}$ and coherent state~$\vert\alpha_1\rangle$~(b)  purity of $\rho_p$. In both panels, $\asymmetry/\qubit = 0.1$ (solid), $\asymmetry/\qubit = 0.2$ (dashed), and~$\asymmetry/\qubit = 0.4$ (dash-dotted).\label{fig:asymmetric_fidelities}}
\end{figure}
In this regime  the reduced density matrix of the phonon is well represented by the coherent state $|\alpha_1\rangle$. The Uhlmann-Jozsa fidelity between $\rho_\mathrm{b}$ and coherent state $|\alpha_1\rangle$ is shown in \cref{fig:asymmetric_fidelities}(a). The corresponding fidelity increases when increasing $\epsilon$. As a result the phonon state population is well approximated by $\alpha_1^2$, which is non-negligible even when $g< g_\mathrm{c}$. This means that subradiance becomes impossible when $\epsilon \neq 0$, which is consistent with the numerical data shown in \cref{fig:Destruciton_of_Spontaneous_symmetry}(b). The fidelity becomes relatively low around $g\sim g_\mathrm{c}$. Importantly the fidelity already approaches unity when $g/g_\mathrm{c} \approx 2$ for all $\epsilon$ shown in the figure. Due to the high fidelity we find that the purity of the phonon density matrix approaches unity, as shown in Fig.~\ref{fig:asymmetric_fidelities}(b). As a result $\rho_\mathrm{b}$ approaches to a pure state when $\epsilon \neq 0$ and $g>g_\mathrm{c}$. This is in sharp contrast to the resonant TQRM, where $\rho_\mathrm{b}$ is a mixed state in the superradiant regime.

\section{Discussion and conclusion\label{sec:conclusions}}
In this work, we have shown that the TQRM can be realized with a pair of trapped Rydberg ions, in conjunction with laser induced spin-dependent force. In this setting the single-body terms in the Hamiltonian can be controlled by tuning the laser and Paul trap electric field. In typical experiments the detuning, Rabi frequency, and trap frequency can be varied flexibly from \si{\kilo\hertz} to \si{\mega\hertz}. The challenging part is to control the tripartite coupling strength $g$, e.g.\ to probe subradiant and superradiant phases. In the Rydberg ion system, $g/g_\mathrm{c} = V_\mathrm{d}^\prime(l_0)/4\sqrt{2M\Omega}$, which depends on the trap frequency implicitly through the equilibrium positions. The dipolar interaction profile is realized by using the microwave dressing scheme discussed in Ref.~\cite{Gambetta_2021}. In this method, coherent coupling of multiple Rydberg states of different parities can shift the dipolar interaction globally, such that the interaction strength is zero at the two-ion equilibrium distance, leading to strong potential gradients that couple the two ionic spins and the phonon mode. For example, we can use same parameters given in Ref.~\cite{Gambetta_2021} to couple states $|50S\rangle$ and $|50P\rangle$ of \ce{^88Sr+} ions. Considering trap frequency $\nu=2\pi\times \SI{1}{\mega\hertz}$, and $\Omega = 2\pi\times \SI{25}{\kilo\hertz}$, we obtain $g/g_\mathrm{c}\approx 0.13$, which falls in the weak coupling (subradiant) regime. To access the strong coupling (superradiant) regime one can choose higher Rydberg states with $n=80$, and tighter trap with $\nu=2\pi\times \SI{2.02}{\mega\hertz}$, which leads to $g/g_\mathrm{c} \approx 1.04$. Hence the tunable laser and trap parameters allows us to explore the physics of TQRM, as well as strong and collective spin-phonon coupled dynamics.

In conclusion, we have studied a novel TQRM where the phonon interacts with two spins simultaneously through the tripartite coupling. The Braak $G$-function is derived analytically for the TQRM, which determines the regular eigenspectra of the model. We have analyzed groundstate properties of the TQRM by varying the detuning $\epsilon$. In the resonant case ($\epsilon=0$), the TQRM can be reduced to the conventional QRM. In the subradiant to superradiant transition region, the phonon becomes a non-Gaussian state whose Wigner function exhibits negative values. In the strong coupling (superradiant) regime, it is found that the reduced density matrix of the phonon is a classical mixture of two coherent states $\lvert \pm \alpha_0\rangle$ with equal probability. In the detuned TQRM ($\epsilon \neq 0$), the subradiance disappears for finite $g>0$. When $g\gg g_c$, we find that $\rho_\text{b}$ becomes a pure state $\lvert \alpha_1\rangle$. This work opens a new way to explore exotic physics through multi-partite couplings between spins and bosons.  One can extend the setting to a longer ion chain, where multiple spins can couple to multiple phonon modes through the tripartite coupling~\cite{wilkinson_spectral_2023}. Such tripartite coupling permits to create, e.g. hyperentanglement in the spin and phonon degrees of freedom simultaneously, which finds quantum information processing  applications ~\cite{renHyperparallelPhotonicQuantum2014,renHighlyEfficientHyperentanglement2015,grahamSuperdenseTeleportationUsing2015,dengQuantumHyperentanglementIts2017,huBeatingChannelCapacity2018,huLongDistanceEntanglementPurification2021a,schollErasurecoolingControlHyperentanglement2023a}.

\begin{acknowledgments}
We thank Tianyi Yan, Lu Qin, Ivo Straka, Harry Parke and Wilson~S.\ Martins for helpful discussions. TJH and WL acknowledge support from the EPSRC through Grant No.\ EP/W015641/1 and the University of Nottingham. IL acknowledges funding from the European Union’s Horizon Europe research and innovation program under Grant Agreement No. 101046968 (BRISQ). This work was supported by the University of Nottingham and the University of T\"{u}bingen’s funding as part of the Excellence Strategy of the German Federal and State Governments, in close collaboration with the University of Nottingham. This work is partially funded by the Going
Global Partnerships Programme of the British Council (Contract No.~IND/CONT/G/22-23/26). The~data used to create the figures in this article can be accessed via an \href{https://doi.org/10.5281/zenodo.10409685}{online repository}.
\end{acknowledgments}
\appendix*
\section{Site-and spin-dependent Stark shift}
The site- and state-dependent Stark shift can be
realized through applying standing wave laser light, which have been extensively studied in the quantum simulation of spin models with trapped
ions. We will provide details of the realization
based on the scheme in Ref.~\cite{PhysRevLett.92.207901}. In the main text, the gradient of the dipolar interaction induces coupling between the two
spins and the motion $H_\mathrm{d} = g\big(\sigma_{1}^{z}\sigma_{1}^{z} + \sigma_{1}^{z} + \sigma_{2}^{z} + 1)(a + a^{\dagger}\big)$.
In the following we show how to  achieve site- and spin-dependent force $F_{j} = - g\big(\sigma_{j}^{z} + \frac{1}{2}\big)\big(a + a^{\dagger}\big)$ to leave only the tripartite coupling. 

\begin{figure}[tb!]
\includegraphics[width=0.8\linewidth]{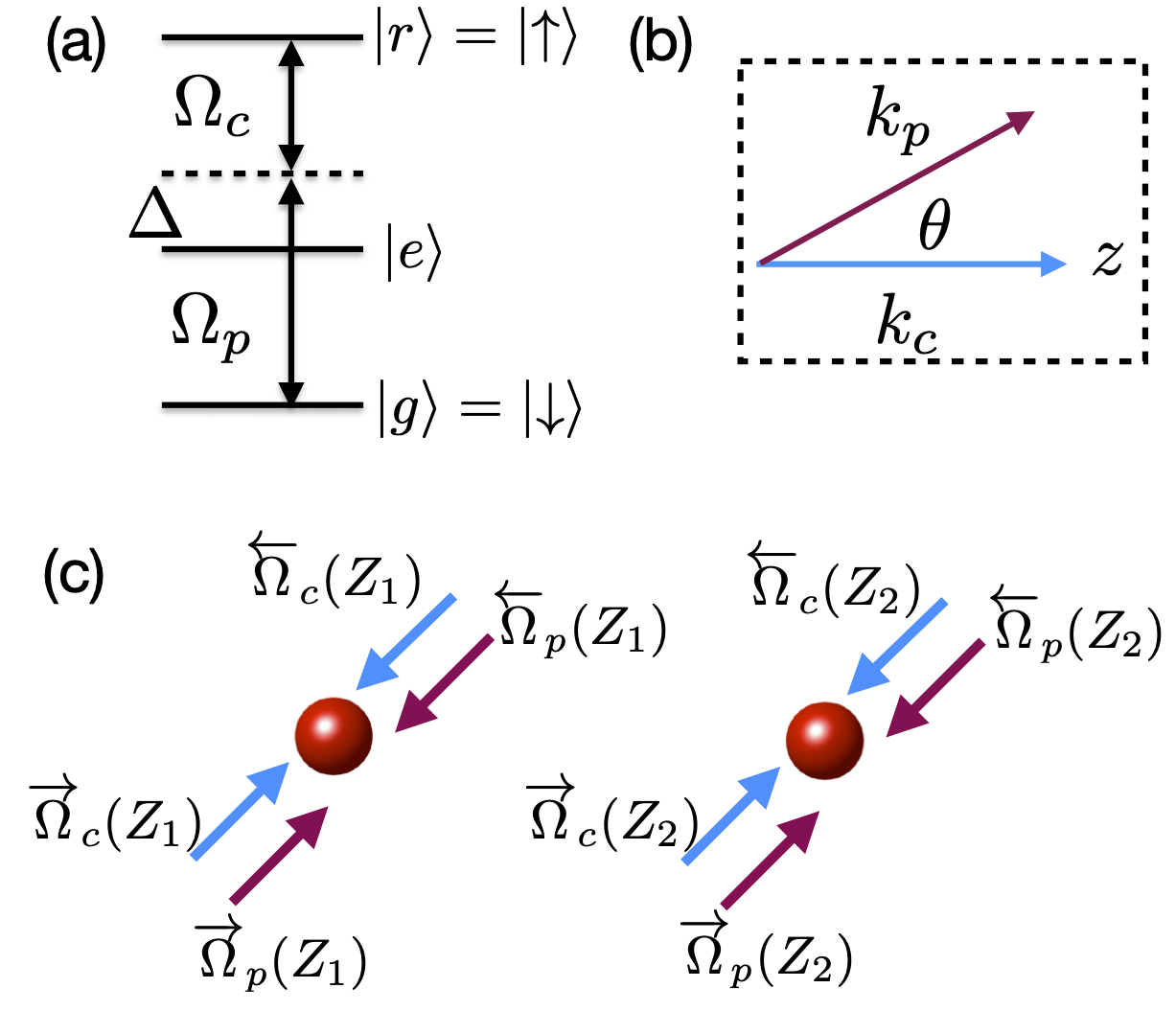}
	\caption{(a)~Level scheme. A probe light and coupling light couple the ground state to the Rydberg state resonantly via an intermediate state. (b) The relative angle between the propagation direction of the probe and coupling light. (c) Both the coupling and probe light are standing waves. The Stark shift will lead to phonon-spin coupling. By tuning the relative phase of the probe light, the spin-dependent force is realized in the Lamb-Dicke regime.~\label{afig:standingwave}}
\end{figure}

Besides the Rydberg excitation laser, we apply another
group of standing wave lasers which drive the ions from the groundstate $\lvert\downarrow\rangle=\lvert g\rangle$
to Rydberg state $\lvert\uparrow\rangle=\lvert r\rangle$ via an intermediate state $|e\rangle$ by probe and coupling
fields. The level scheme is depicted in Fig.~\ref{afig:standingwave}(a).   In case of \ce{Sr+}
ions, the wave length of the probe and coupling light are
$\lambda_\mathrm{p} = \SI{243}{\nano\meter}$ and $\lambda_\mathrm{c} = \SI{243}{\nano\meter}$. For concreteness,
we assume the coupling light propagates along the trap z-axis (or a
small angle with respect to the trap axis). The propagation direction of
probe light has an angle $\theta = \arccos(k_\mathrm{c}/k_\mathrm{p})$ with
$k_\mathrm{p} = 2\pi/\lambda_\mathrm{p}$ and $k_\mathrm{c} = 2\pi/\lambda_\mathrm{c}$, as shown
 Fig.~\ref{afig:standingwave}(b). This will ensure that the wavevectors of both fields along the trap axis are identical to ${k = k}_\mathrm{c}$. The probe and coupling
fields (along the trap axis) of the $j$th ion are given by,
\begin{eqnarray}
	\Omega_\mathrm{p}^{(j)} = {\overline{\Omega}}_\mathrm{p}\big( \ex{\mathrm{i}kz_{j}} + \ex{-\mathrm{i}kz_{j} +\mathrm{i}\varphi_{j}} \big)\text{,}\nonumber\\
	\Omega_\mathrm{c}^{(j)} = {\overline{\Omega}}_\mathrm{c}\big( \ex{\mathrm{i}kz_{j}} + \ex{- \mathrm{i}kz_{j} + \mathrm{i}\beta_{j}}\big)\text{,}\nonumber
\end{eqnarray}
where ${\overline{\Omega}}_\mathrm{p}$ and ${\overline{\Omega}}_\mathrm{c}$ are
amplitudes of the Rabi frequency applied on the $j$th ion. $\varphi_{j}$ and $\beta_j$ are the relative phase of the probe and coupling light. $z_j$ is the small deviation from the equilibrium position. The Hamiltonian describes this interaction reads,
\begin{equation}
\label{eq:a1}
H^{(j)} = \Delta\sigma_{ee}^{(j)} + \left(\Omega_\mathrm{c}^{(j)}\sigma_{er}^{(j)} + \Omega_\mathrm{p}^{(j)}\sigma_{ge}^{(j)} + \text{H.c.} \right),
\end{equation}
with $\sigma_{ab}^{(j)}=\lvert a\rangle\langle b\rvert$ for the $j$th ion.
Assuming both fields are weak and $\Delta \gg \Omega_\mathrm{c}^{(j)},\Omega_\mathrm{p}^{(j)}$,
state $\lvert e \rangle$ is adiabatically eliminated. We obtain an effective Hamiltonian between state $\lvert g\rangle$ and $\lvert r\rangle$
\begin{equation}
	H_\mathrm{e}^{(j)} = -\frac{\big\vert\Omega_\mathrm{c}^{(j)}\big\vert^2}{\Delta}\sigma_{rr}^{(j)} - \frac{\big\vert\Omega_\mathrm{p}^{(j)}\big\vert^2}{\Delta}\sigma_{gg}^{(j)} - \left(\frac{\Omega_\mathrm{p}^{(j)}\Omega_\mathrm{c}^{(j)}}{\Delta}\sigma_{rg}^{(j)} + \text{H.c.}\right)\text{.}
\end{equation}

Following the scheme of \citet{PhysRevLett.92.207901}, we now show how to realize the site-dependent force. In our setting,  the relative phase of the standing wave fields will be different 
for the first and second ion, while other parameters are same. 
We will set the phase of the first ion to be $\varphi_{1} = {- \beta}_{1} = -\pi/2$. The Rabi
frequencies are given by
\begin{align*}
	\Omega_\mathrm{p}^{(1)} & = \overline{\Omega}_\mathrm{p}\big(\ex{\mathrm{i}kz_1} - \mathrm{i}\ex{-\mathrm{i}kz_1}\big)\text{,} \\
	\Omega_\mathrm{c}^{(1)} & = \overline{\Omega}_\mathrm{c}\big(\ex{\mathrm{i}kz_1} + \mathrm{i}\ex{-\mathrm{i}kz_1}\big)
\end{align*}

In the Lamb-Dicke regime, we  Taylor expand the ionic
coordinate around the equilibrium position. The state dependent Stark
shifts are obtained,
\begin{eqnarray}
\frac{\big| \Omega_\mathrm{p}^{(1)} \big|^{2}}{{\overline{\Omega}}_\mathrm{p}^{2}} &=& 2 - 2\sin\left( 2kz_{1} \right)\nonumber\\
& \approx& 2 - \ 2\sqrt{2}\eta\big( a + a^{\dagger} \big) - {2\sqrt{2}\eta}_{\text{c.m.}}\left( a_{\text{c.m.}} + a_{\text{c.m.}}^{\dagger} \right), \nonumber\\
\frac{\big| \Omega_\mathrm{c}^{(1)} \big|^{2}}{{\overline{\Omega}}_\mathrm{c}^{2}} &=& 2 + 2\sin\left( 2kz_{1} \right) \nonumber\\
&\approx& 2+2\sqrt{2}\eta\big( a + a^{\dagger} \big) + {2\sqrt{2}\eta}_{c.m.}\big( a_{\text{c.m.}} + a_{\text{c.m.}}^{\dagger} \big),\nonumber\\
\frac{\Omega_\mathrm{p}^{(1)}\Omega_\mathrm{c}^{(1)}}{{\overline{\Omega}}_\mathrm{p}{\overline{\Omega}}_\mathrm{c}} &=& 2\cos\left( 2kz_{1} \right)\approx 2,\nonumber
\end{eqnarray}
where $\eta$ and $\eta_{\text{c.m.}}$ are the Lamb-Dicke parameter of the
breathing and c.m. modes. In the above equations, we have considered the
linear order terms as allowed in the Lamb-Dicke regime.

In this regime, the site-dependent Stark shift $F_{1}$ considered in the main text is realized when the Rabi
frequencies~$\overline{\Omega}_\mathrm{p}^{2} = \frac{\Delta}{4\sqrt{2}\eta}g$, and~${\overline{\Omega}}_\mathrm{c}^{2} = \frac{3\Delta}{4\sqrt{2}\eta}g$.
The effective Hamiltonian of the first ion becomes
\begin{flalign}\label{eq:ion1}
	H_\mathrm{e}^{(1)} & = -g\big(a + a^\dagger\big)\left(\sigma_1^z + \frac{1}{2}\right) - g\sigma_1^x - \frac{g}{2\sqrt{2}\eta}\big(\sigma_1^z + 2\big) \nonumber \\
	& \hphantom{=} + g\sqrt[4]{3}\big(a_\mathrm{c.m.} + a_\mathrm{c.m.}^\dagger\big)\left(\sigma_1^z + \frac{1}{2}\right)\text{.}
\end{flalign}

The phases of the second ion are  $\varphi_{1} = {- \beta}_{1} = \frac{\pi}{2}$. This choice of the phase is necessary as the breathing mode vector has opposite signs for the two ions. With this phase,
one carries out same calculations and obtains the Hamiltonian of the second ion
\begin{flalign}\label{eq:ion2}
	H_\mathrm{e}^{(2)} & = -g\big(a + a^\dagger\big)\left(\sigma_2^z + \frac{1}{2}\right) - g\sigma_2^x - \frac{g}{2\sqrt{2}\eta}\big(\sigma_2^z + 2\big) \nonumber \\
	& \hphantom{=} + g\sqrt[4]{3}\big(a_\mathrm{c.m.} + a_\mathrm{c.m.}^\dagger\big)\left(\sigma_2^z + \frac{1}{2}\right)\text{.}
\end{flalign}

Combining \cref{eq:ion1,eq:ion2} the total Hamiltonian becomes
\begin{flalign}\label{eq:coupling}
	H_\mathrm{e} & = -g\big(a + a^\dagger\big)\big(\sigma_1^z + 1\big) - g\big(\sigma_1^z + \sigma_2^x\big) \nonumber \\
	& \hphantom{=} - \frac{g}{2\sqrt{2}\eta}\big(\sigma_1^z + \sigma_2^z + 4\big) + g\sqrt[4]{3}\big(a_\mathrm{c.m.} + a_\mathrm{c.m.}^\dagger\big)\big(\sigma_2^z - \sigma_1^z\big)\text{.}
\end{flalign}
On the right hand side of the Hamiltonian (\ref{eq:coupling}), the first term
gives the required tripartite coupling. The
second and third term will contribute to the overall single-spin terms in
 Hamiltonian (\ref{eq:EQRM}), where the constant term will not affect the
dynamics. The last term, on the other hand, should be eliminated, as it
could mix different symmetry sectors. This can be achieved by cooling
 the c.m. mode in the zero-phonon state
$\lvert 0\rangle_{\text{c.m.}}$~\cite{PhysRevLett.62.403}, (or other Fock state
$\lvert  n \rangle_{\text{c.m.}}$) such that
$\langle 0\rvert\big(a_{\text{c.m.}} + a_{\text{c.m.}}^{\dagger} \big)\lvert 0 \rangle_{\text{c.m.}} = 0$. Alternatively, one could apply the dynamical decoupling scheme~\cite{barthel_robust_2023} to decouple the c.m. mode from the dynamics. 
As a result, the tripartite coupling can be achieved.

\end{document}